\documentclass{aa} 
\usepackage{graphicx}
\usepackage{longtable}
\usepackage{lscape}
\usepackage{amsmath}
\usepackage{lmodern}

\begin{document} 

\title{Relationship between Ca and H$\alpha$ chromospheric emission in F-G-K stars: indication of stellar filaments?}

    \titlerunning{Relationship between Ca and H$\alpha$ chromospheric emission in F-G-K stars}

\author{N. Meunier \inst{1}, M. Kretzschmar \inst{2}, R. Gravet  \inst{2}, L. Mignon \inst{1},  X Delfosse \inst{1} 
  }
\authorrunning{Meunier et al.}

\institute{
Univ. Grenoble Alpes, CNRS, IPAG, F-38000 Grenoble, France \email{nadege.meunier@univ-grenoble-alpes.fr} \\
\and
LPC2E, University of Orl\'eans, CNRS, 3A avenue de la Recherche Scientifique, 45071 Orl\'eans Cedex 2, France
}

   \date{}

\offprints{N. Meunier}

\date{Received ; Accepted}

\abstract{
Different relationships between the H$\alpha$ and Ca II chromospheric emissions have been reported in solar-type stars. In particular, the time-series of emissions in these two lines are clearly anti-correlated for  a few percent of the stars, contrary to what is observed on the Sun.}  
{Our objective is to characterise these relationships in more detail using complementary criteria, and to constrain the properties of filaments and plages that are necessary to explain the observations.}
{We analysed the average level and variability of the H$\alpha$ and Ca II emission for 441 F-G-K stars, paying particular attention to their (anti-)correlations on both short and long timescales. We also computed synthetic H$\alpha$ and Ca II time-series for different assumptions of plage and filament properties and compared them with the observations. }
{We were not able to find plage properties that, alone,  are sufficient to reproduce the observations at all timescales simultaneously, even when allowing different H$\alpha$ and Ca II emission relationships for different stars. We also specified the complex and surprising relationship between the average activity levels of both indexes, in particular for low-activity stars.}
{We conclude that plages alone are unlikely to explain the observed variety of relationships between Ca II and H$\alpha$ emission, and that the presence of other phenomena like filaments may help to reconcile the models with observations. 
}


   \maketitle
%



\section{Introduction}

Chromospheric emission is a widely used indicator of stellar activity, especially based on the emission in the core of Ca II H and K lines \cite[e.g.][]{baliunas95,radick98,lockwood07,hall09,radick18}. It is also of great interest in other fields, as the $\log R'_{\rm HK}$  index  is often used as a proxy to disentangle the exoplanet signal from the stellar activity contribution in radial velocity time-series 
\cite[e.g.][]{saar97,boisse09,lagrange10b,meunier10a,pont11,dumusque12,robertson14,rajpaul15,lanza16,diaz16,dumusque16,dumusque17,borgniet17}. This is particularly useful because the long-term variability of stars like the Sun is strongly affected by the inhibition of the convective blueshift in plages \cite[][]{saar97,meunier10a}, although some complex behaviour must be taken into account in order to be able to reach very low mass planets \cite[][]{meunier13,meunier19c}. 
In addition to its effect on exoplanet detectability, stellar activity may affect exoplanet habitability, and the stellar wind can erode the atmosphere of terrestrial-type planets, causing a significant loss of volatiles (including H$_{\rm 2}$O) and influencing atmospheric evolution \citep[see e.g.][]{vidotto2013}. 
Therefore, a good understanding of the processes producing stellar variability is important.

Several studies have compared the variability of stellar chromospheric emission by analysing different chromospheric indexes and in particular the Ca II H and K and H$\alpha$ indexes. The first studies of this kind \cite[][]{giampapa89,strassmeier90,robinson90,pasquini91} were limited to using non-simultaneous observations and found that chromospheric indicators are well correlated with each other  most of the time \cite[see the review by][]{linsky17}, similarly to the very good correlation observed on the Sun \cite[][]{livingston07}. This is easily explained by the high level of correspondence between the disk-resolved images in both lines, which are dominated by plages. \cite{maldonado19} observed an anti-correlation when observing the Sun with HARPS-N, which nevertheless remains to be understood. \cite{marchenko21} also found that the other Balmer lines, such as H$\beta$, H$\gamma$, and H$\delta,$ also do not trace the usual chromospheric activity indicators  as well as expected on rotational timescales, possibly being more affected by photospheric variability due to spots. 
This was confirmed by \cite{cin07} who used simultaneous observations and a better temporal sampling, and found a correlation between averaged (overall values for each star) indexes in a large sample of F-G-K stars. However, these latter authors also obtained a puzzling result when analysing individual time-series: a few percent of the  stars in their sample exhibit an anti-correlation between the Ca II H and K index and the H$\alpha$ index. This was confirmed on a larger sample by  \cite{gomes14} for stars with similar spectral types. Such stars therefore have a behaviour that is very different from that of the Sun, and this  remains to be understood.

\cite{meunier09a} proposed that this behaviour could be caused by a larger number of (or bigger) filaments in those stars: their presence would affect the correlation between indexes by adding a strong absorption in H$\alpha$ that may in some cases compensate the plage signal. However, this could also be caused by a peculiar behaviour of the H$\alpha$ emission in plages at low activity regimes, as has been proposed by \cite{cram79}  for M stars: when the activity increases (as is the Ca II H and K index), the H$\alpha$ line first exhibits a stronger absorption, before going into emission at higher  levels of activity. The increase in absorption could be due to an increase in temperature gradient \cite[][]{cram79} or to a stronger density of the plasma in the chromosphere \cite[][]{cram85} leading to a dominant effect of collisions: both processes are considered in \cite{cram87}. However, this has only been proposed for M stars and it is therefore speculative to apply this mechanism to F-G-K stars.

Our objective here is to implement additional diagnoses to describe and explain those observations, and in particular to distinguish between these two proposed mechanisms. We considered a less selective and larger sample of stars than \cite{gomes14} in order to present a broader picture: we included  more active stars, as well as a few  subgiants. A more sophisticated approach is also necessary because, as shown below, the anti-correlations found by \cite{cin07} are not the only puzzling result: there are a large number of stars with a correlation close to zero despite significant long-term variability in Ca II (we see in the following that 40\% of the stars with very low correlation and good temporal sampling are in this configuration). We specifically aim to investigate whether plages and/or filaments can explain these unexpected (null and negative) observed correlations by attempting to answer two major questions: {Can the unexpected correlations be explained by a complex relationship between Ca and H$\alpha$ emission in plages in some activity regimes? {Otherwise, is the presence of filaments necessary to explain the observations?} To this end, we implemented a statistical analysis on our large sample and also analysed stars with good temporal sampling and individual cases in more detail.

The outline of the paper is as follows: In Sect.~2, we describe the star sample and its properties, and the derivation of the chromospheric indexes.  The relation between average Ca II and H$\alpha$ emissions is studied in Sect.~3. In Sect.~4, we describe the tools used to characterise the variability of the stars at different timescales and how this variability relates to stellar parameters. 
In Sect.~5, we specifically investigate the possible role of plages by comparing the expected (modelled) correlations and amplitudes of H$\alpha$ based on the Ca II `plage-induced' variability to the observed values.  Finally, we discuss a possible interpretation in terms of filaments in Sect.~6 and conclude in Sect.~7.  \\

\section{Sample and chromospheric activity indexes}

In this section, we first describe the stellar sample, followed by the computation of the Ca II H and K and   H$\alpha$ indexes. We present the selection procedure.  

\begin{figure}
\includegraphics{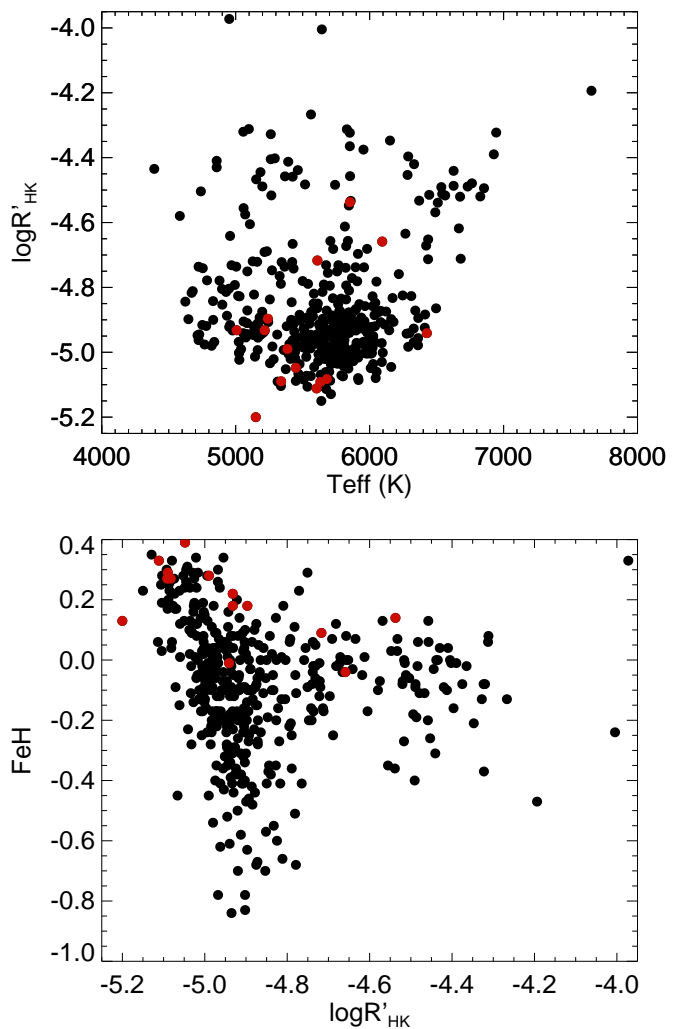}
\caption{
{\it Upper panel:} $\log R'_{HK}$ versus T$_{\rm eff}$ for our sample of 441 stars. Subgiants (identified with a luminosity class of IV in SIMBAD/CDS) are shown in red. 
{\it Lower panel:} Same but for the metallicity versus $\log R'_{HK}$. 
}
\label{sample}
\end{figure}

\begin{figure}
\includegraphics{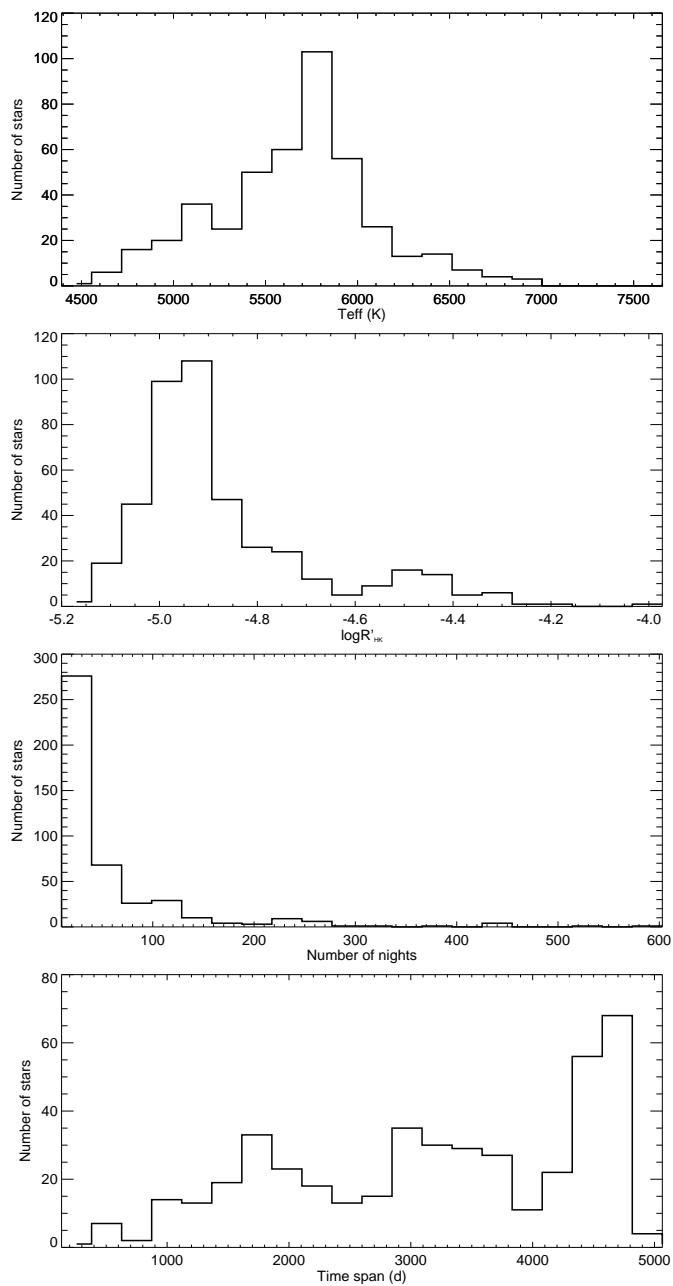}
\caption{
{\it First  panel:} Distribution of  T$_{\rm eff}$ values for our sample of 441 stars. 
{\it Second panel:} Same but for $\log R'_{HK}$.
{\it Third panel:} Same but for the number of nights. 
{\it Fourth panel:} Same but for the time span. 
}
\label{sample_dist}
\end{figure}

\subsection{Stellar sample}

We  gathered HARPS (High Accuracy Radial velocity Planet Searcher) data for a large sample of F-G-K stars, covering spectral types from F0 to K6. The spectra were retrieved from the European Southern Observatory (ESO) archive. These stars have been observed and studied in various surveys and in particular by \cite{sousa08}, \cite{ramirez14}, \cite{marsden14}, \cite{datson14}, \cite{borgniet17}, \cite{lagrange13}, and \cite{gray15}. We have already studied most of these stars in \cite{meunier17b} to estimate the amplitude of their convective blueshift and its dependence on the activity level.  In the present paper, our primary sample includes 465 stars (selected from published surveys), of which we have selected stars with observations on  at least ten different nights after the selection process (see Sect.~2.2), resulting in a final sample of 441 stars. Most of these are old main sequence stars, with a strong bias towards solar-type and quiet stars (and therefore old stars), as shown in Fig.~\ref{sample}. Stars with a low  $\log R'_{HK}$ also tend to have a higher metallicity, as already observed in previous works \cite[][]{jenkins08,meunier17b}, without a clearly identified explanation so far. The distributions in T$_{\rm eff}$ and $\log R'_{HK}$ are shown in Fig.~\ref{sample_dist}. In our sample, G stars represent the largest fraction (269 stars), followed by K stars (99 stars) and F stars (73 stars). Furthermore, the sample  includes young main sequence active stars (54 have an average $\log R'_{HK}$ above  -4.6), as well as a few subgiants and one giant, as summarised in Table~\ref{tab_classe}.
Table~\ref{tab_sample} lists all 441 stars and their fundamental parameters.

HARPS has been operating since 2003, allowing very good temporal coverage of those stars. We analysed data obtained between 2003 and 2017.
The median time-span coverage is 9.1~yr. Each star has observations over a time-span coverage of between 133 and 5064 nights. A total of  26579 nights of observations are analysed. The distributions of the number of nights  per star and of the temporal coverage are shown in Fig.~\ref{sample_dist}. A total of 429 stars, which is more than 97\% of the sample, have a time-span longer than 1000 days. The number of independent  nights with observations per star ranges from 10 (after selection, see below) to 603, with a median of 32 nights. Of these 429 stars, 137 have been observed for more than 50 nights, and 70 stars  for more than 100 nights.

\begin{table}
\caption{Stellar type statistics}
\label{tab_classe}
\begin{center}
\renewcommand{\footnoterule}{}  
\begin{tabular}{ll}
\hline
Luminosity class  & Number  \\
\hline
V & 415  \\
IV & 13  \\
IV-V & 9  \\
III & 1 \\
unidentified & 3  \\
\hline
\end{tabular}
\end{center}
\tablefoot{Luminosity classes are from the CDS, except for HD163441 \cite[][]{lorenzo18}, HD78534 \cite[][]{dossantos16}, and HD141943 \cite[][]{grandjean20}.
}
\end{table}

\subsection{Computation of activity indexes}

We computed the S-index, hereafter S$_{\rm Ca}$,   in a classical way to anchor our measurements to the historical values of the Mount Wilson $R'_{HK}$, as in previous works \cite[e.g.][]{baliunas95,cin07,meunier17b}. We took the ratio between the flux integrated in the core of the Ca II H and K lines (using a triangular weighting function with a width of 1.09 $\AA$) and the continuum estimated between 3891 and 3911 $\AA$ plus that between 3991 and 4011 $\AA$. The variability analysis was performed on this indicator but we also computed the usual $\log R'_{HK}$, using the photospheric contribution subtraction  and the correction by the bolometric flux from \cite{noyes84}, to derive average activity levels that can be compared for stars of different spectral types. 

We computed the H$\alpha$ index, hereafter S$_{\rm H\alpha}$ as in \cite{gomes11}: we integrated the flux in the core of the H$\alpha$ line over a bandwidth of 1.6 $\AA$, which was then normalised by the flux in the continuum in the 6545--6556 $\AA$ and 6576--6585 $\AA$ domains. Again, this indicator was used in our temporal variability analysis of each star and therefore does not need to be calibrated in absolute flux. 
The average flux $I_{H\alpha}$ was also computed for complementary analysis after correction of a colour-dependence (B-V), described in Appendix B: we then considered the log of this corrected index.

The formal uncertainties on each value, computed from the propagation of the uncertainties due to the photon noise,  are very small, and we chose to adopt a more conservative approach  in this work, as systematic sources of uncertainties might be present. We first selected all nights with at least 50 observations of the same star, which lead us to a sample of 133 nights for 17 stars.  We then computed the standard deviation of the flux values over each night, and plotted it as a function of signal-to-noise ratio (S/N) in the Ca II H and K orders, or in the H$\alpha$ order, respectively, for the two indexes. We assumed that the stars with the smallest dispersion have no intrinsic variability over one night, and that the dispersion is due to the uncertainty on the measurements only. 
This would not take any long-term systematic error into account, which we cannot estimate.
The lower envelope  (which is higher than the formal errors from the photon noise) provided us an estimate of the uncertainty as a function of S/N on the spectra \cite[see also][for a discussion about the instrumental noise floor]{lovis11b}. We obtained the following  uncertainties: (1) that on  S$_{\rm Ca}$  (S/N considered for the seventh order of the HARPS spectra, i.e. for Ca II K ):
\begin{equation}
{\rm if} \; S/N>40, \sigma_{\rm Ca}=\sigma_0=0.0005, 
\end{equation}
\begin{equation}
{\rm if} \;  S/N<40, \sigma_{\rm Ca}=\frac{\sigma_0+(S/N-40)^2}{2\; 10^5} 
,\end{equation}
and (2) that on S$_{\rm H\alpha}$ (S/N considered for the sixty-seventh order of the HARPS spectra):
\begin{equation}
{\rm if}  \; S/N>40, \sigma_{\rm H\alpha}=\sigma_0'=0.0003 
,\end{equation}
\begin{equation}
{\rm if}  \; S/N<40, \sigma_{\rm H\alpha}=\frac{\sigma_0'+(S/N-300)^2}{5\; 10^7} 
.\end{equation}
As some of the spectra have a very low S/N, we implemented a selection based on a threshold adapted to each star in order to retrieve as many observations as possible. For each star, we analysed all individual measurements of the Ca II index as a function of S/N. The same was done for the H$\alpha$ index. 
For different bins in S/N (of width 10), we computed the average and the standard deviation of the values:
in principle, S/N bins with a larger dispersion than bins with a very good S/N should be discarded, as should those with an average far from the other values. A reference level and standard deviation were first  estimated for all points with S/N$>$30 (m$_{\rm 30}$ and $\sigma_{\rm 30}$ respectively). We identified the first S/N bin for which the standard deviation was lower than $\sigma_{\rm 30}$ and the first S/N bin for which the average was smaller than m$_{\rm 30}$+3$\sigma_{\rm 30}$, and took the highest of these S/N values. We kept all points with a S/N higher than this final threshold. In addition, we kept all points when there were more than 15 measurements during a given night, because even if the S/N is smaller the average should be statistically significant. Out of the original 88324 individual observations (for 465 stars), a total of 37224 measurements are kept. After binning over each  night, this leads to 26672 individual nights. Finally, we kept stars with observations on at least ten nights, giving a total of 441 stars representing a total of 26579 nights for our final sample.

\begin{figure}
\includegraphics{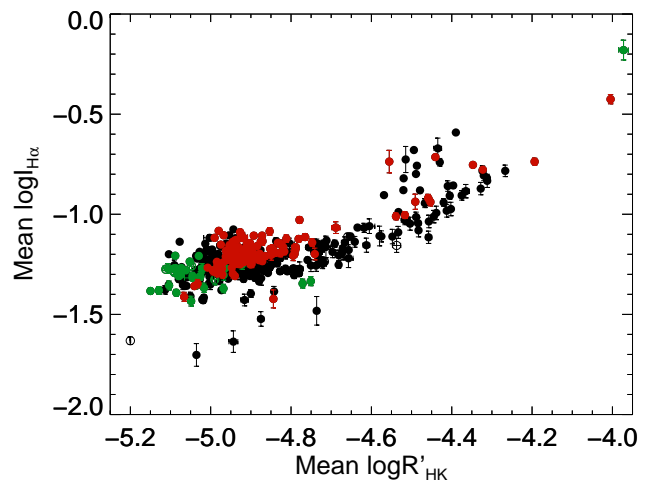}
\caption{
Average $\log I_{H\alpha}$ versus average $\log R'_{HK}$ for each star. Stars with metallicity lower than -0.2 are in red, and larger than 0.2 in green, while stars with metallicity between -0.2 and 0.2 are in black. 
}
\label{moy}
\end{figure}

\begin{figure*}
\includegraphics{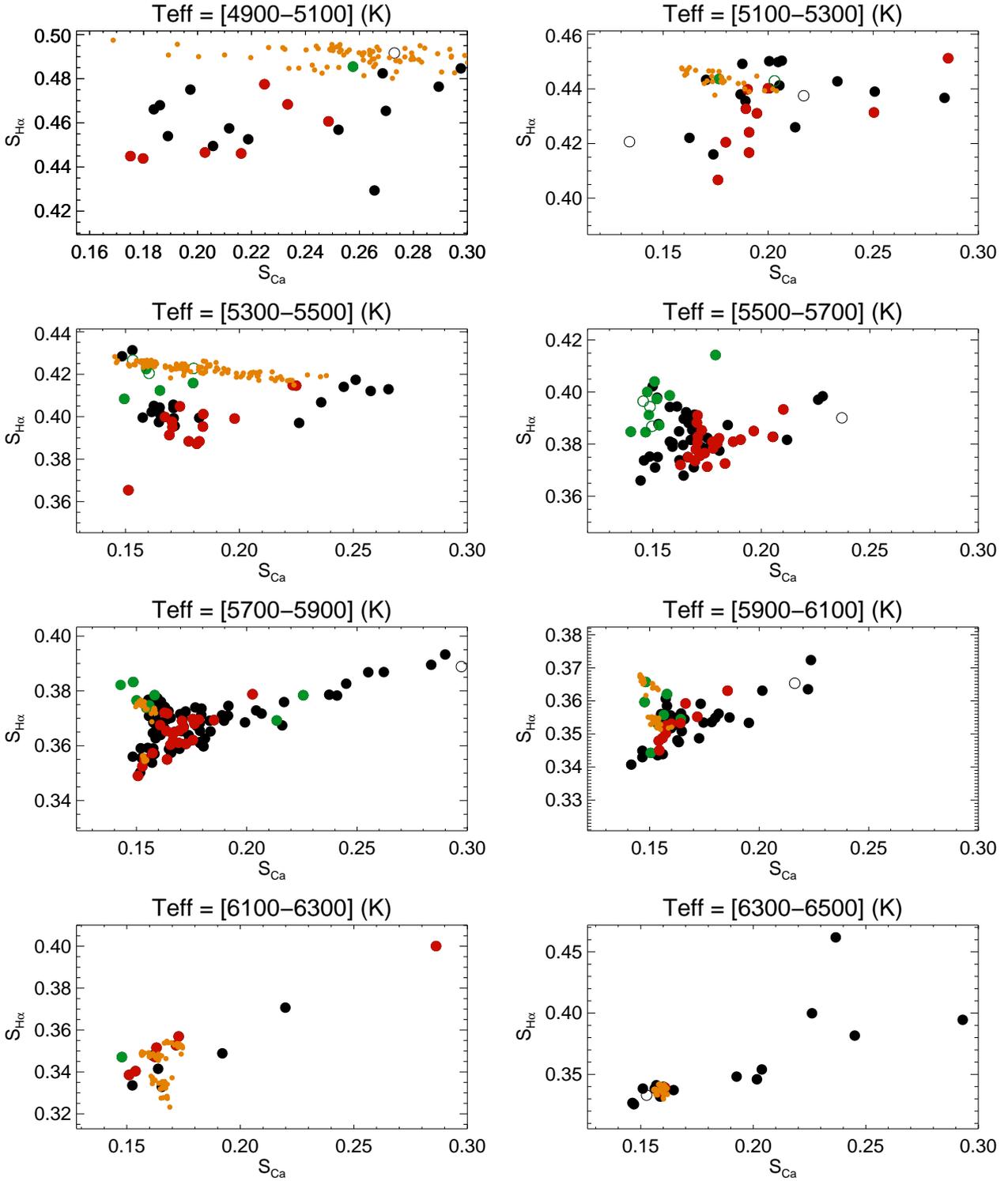}
\caption{
Average S$_{\rm H\alpha}$ versus average S$_{\rm Ca}$ for quiet stars. Stars with metallicity lower than -0.2 are in red, stars with metallicity between -0.2 and 0.2 are in black, and stars with metallicity larger than 0.2 are in green. Open symbols are subgiants. The small orange dots are nightly observations for stars with a correlation below -0.5. Each panel corresponds to a different T$_{\rm eff}$ range. 
}
\label{moyzoom}
\end{figure*}

\section{Average chromospheric indexes}

Before analysing the temporal variability, we consider the relation between the average emission levels. Figure~\ref{moy} shows the average $\log I_{H\alpha}$ of all stars in our sample versus their average $\log R'_{HK}$. We observe a good correlation between the two indexes (Pearson coefficient of 0.82). However, there is significant dispersion, especially for quiet stars: the correlation for stars with a $\log R'_{HK}$ of lower than -4.6 \cite[as in e.g.][]{gomes14} is for example much lower (0.44). This is very similar to what has been found in previous works for similar stars: \cite{cin07} found a strong dispersion, as did \cite{gomes14} with stars with a $\log R'_{HK}$ mostly below -4.6.
A few stars seem to depart from the general behaviour: a few low-activity stars have a lower H$\alpha$ emission (stars with $\log I_{H\alpha}$ below $\sim$-1.5 in the lower left part of Fig.~\ref{moy}), and a few active stars have a larger H$\alpha$ emission than most stars with the same average $\log R'_{HK}$ (they have $\log R'_{HK}$ around -4.5 and $\log I_{H\alpha}$ above -0.8 in Fig.~\ref{moy}). Such stars are either low-T$_{\rm eff}$ or high-T$_{\rm eff}$ stars, and this effect is likely to be caused by a residual of the correction applied in Appendix B, because of the low number of stars in these regimes. This is indeed apparent in Fig.~\ref{corr_halpha}.

Even though the indexes have been corrected for the photospheric contribution and, in the case of $\log R'_{HK}$, by the bolometric flux, the range of spectral types and classes is such that it is useful to investigate this flux--flux relationship by considering small bins of T$_{\rm eff}$ independently. The results are shown in Fig.~\ref{moyzoom}, for  200 K-wide bins, for the  S$_{\rm H\alpha}$ and S$_{\rm Ca}$ indexes. The extension towards active stars (S$_{\rm Ca}$>0.3) is not shown here for clarity, but exhibits a correlation similar to that of Fig.~\ref{moy} (see Appendix C.2). At these low activity levels, the relationship appears  more complex than a linear correlation. At very low activity levels, the dispersion in the H$\alpha$ index increases, with the possible existence of two branches, which are particularly clearly seen between 5100 K and 6100 K where a large number of stars are available: a lower branch that corresponds to an overall monotonous relation between the Ca II and H$\alpha$ indexes, and a upper branch (seen for S$_{\rm Ca}$ typically below $\sim$0.2 and a level in S$_{\rm H\alpha}$ depending on T$_{\rm eff}$)  where the H$\alpha$ emission first appears to decrease with increasing Ca II activity level. We note that high-metallicity stars (in green) are seen predominantly on the higher branch. This may be similar to what was observed for M stars by \cite{scandariato17}, although here we also have stars on the lower branch.

The orange dots in Fig.~\ref{moyzoom} show "daily" values for stars exhibiting anti-correlation in their S$_{\rm H\alpha}$ and S$_{\rm Ca}$ time-series ($c<-0.5$), as detailed in the following section. It is noticeable that these stars are predominantly observed on the upper branch. It is therefore possible that the observed anti-correlation corresponds to a specific behaviour represented by this upper branch, with an increasing average H$\alpha$ index when  the  activity level decreases as represented by the average Ca index, which provides important information for an interpretation of the anti-correlation. 
This is the first time that such a behaviour is seen for F-G-K stars to our knowledge. We identify three possible explanations for this behaviour in the low-activity regime, and consider these further in the following sections:

\begin{itemize}
\item{{\it There are two different relations between Ca II and H$\alpha$ plage emission at low activity level, corresponding to the two observed branches}. \cite{cram79} proposed that the Ca~II--H$\alpha$ relation in M stars was non-monotonous because of the fact that when activity increases, the absorption in H$\alpha$ first increases before switching to emission \cite[see also][]{cram87}, theoretically leading to a U-shape in the Ca II-H$\alpha$ relationship. Attempts to observe this effect in M stars have been made, with a first result obtained by \cite{giampapa89} on a small sample with non-simultaneous observations of Ca II and H$\alpha$. \cite{walkowicz09} and \cite{gomes11} may have observed hints of this effect in a small sample of M stars, although it mostly shows a flat H$\alpha$ emission in the low-activity range with a strong dispersion rather than the expected U-shape. However, this was clearly observed for M stars by \cite{scandariato17}. We cannot exclude that we observe a similar effect  in more massive stars, with emission in plages having a different slope (between Ca II and H$\alpha$) depending on their average H$\alpha$ level, although we have seen no such studies in the literature for solar-type stars. We test the effect of this assumption in Sect.~5 by modelling H$\alpha$ time-series deduced from Ca II time-series using such slopes. The modelling details are described in Appendix C.2. In Appendix C.3, we also investigate whether or not the  different slopes observed for different H$\alpha$ levels could be due to additional processes. 
 }
\item{{\it The two branches are caused by the presence of filaments.} In this case, both  the presence of anti-correlation or almost null correlations and the dispersion could be due to the same effect, namely the presence of filaments strongly affecting the H$\alpha$ time-series, while the Ca II and H$\alpha$ plage emission would be the same for all activity levels. One difficulty with this scenario is that, because filaments lead to H$\alpha$ absorption, we expect the anti-correlated stars to be closer to the lower branch in Fig.~\ref{moyzoom}, which is not the case, meaning that this assumption is probably not sufficient to explain the observation.  }
\item{{\it The two branches are due to a different average H$\alpha$ level of the quiet star, that is, a different basal flux}. This would need to be true only for relatively quiet stars, with a basal flux depending on other stellar parameters (e.g. metallicity). Furthermore, \cite{cram79} indicated that for M stars at least, it is easier to obtain relatively high H$\alpha$ emission for stars with low metallicity, which is the opposite trend of what we observe here. For M stars too, \cite{houdebine97} argued that Ca II emission is also sensitive to metallicity (lower metallicity corresponding to lower activity levels, which is different from what is observed in the lower panel of Fig.~\ref{sample}), while H$\alpha$ emission is not sensitive to it. While this explanation remains a possibility, we do not investigate it further  in this study. }
\end{itemize}


\section{Observed correlations and variability}

In this section, we analyse the time-series obtained in Ca II and H$\alpha$ for our large sample of stars. We start by describing the timescales that we considered and the computation of correlations between time-series, and then establish whether or not the observed variability is significant for the whole sample of stars. We then use this information to test whether those observations can be explained by the presence of plages alone based on three assumptions for the relationship between the Ca II and H$\alpha$ emission in plages.


\subsection{Methods: Correlation between H$\alpha$ and Ca II emission on different temporal scales}

We describe the methods used to analyse the time-series obtained for each star. We detail how the analysis is performed for different  timescales. 

\subsubsection{Computation of the correlations}

\begin{figure*}
\includegraphics{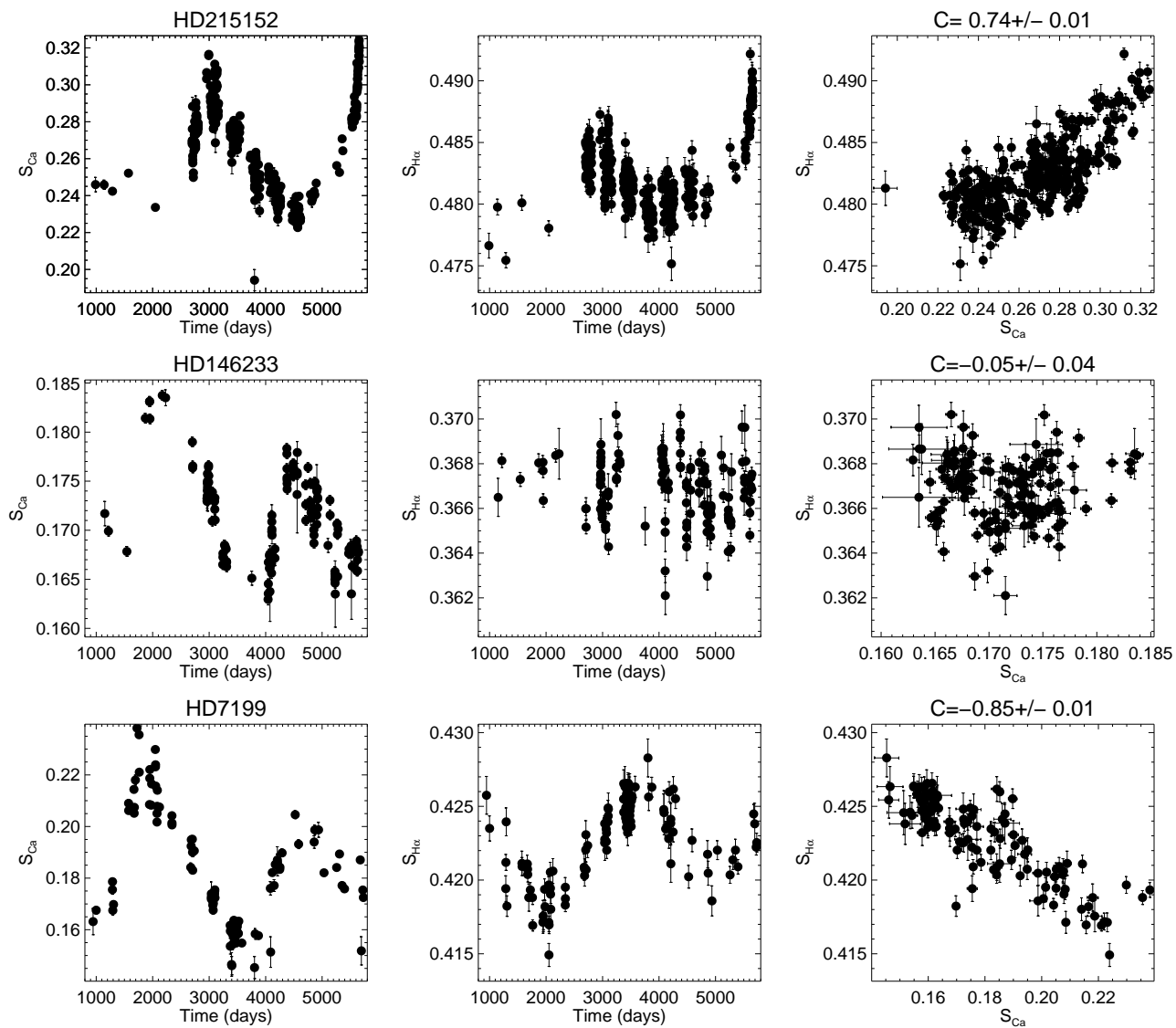}
\caption{
{\it Upper panel:} S$_{\rm Ca}$ versus time (Julian day minus 2452000 days), S$_{\rm H\alpha}$ versus time, and  S$_{\rm H\alpha}$ versus S$_{\rm Ca}$ for HD215152. C is the global correlation between the two indexes.   
{\it Middle panel:} Same for HD146233.
{\it Lower panel:} Same for HD7199. 
}
\label{exemple}
\end{figure*}

Our primary tool to analyse the relationship between  the Ca II and H$\alpha$ emission is the correlation, defined by the linear Pearson coefficient (a comparison with the Spearman coefficient gives very similar values). We first computed this coefficient for each S$_{\rm Ca}$(t) and  S$_{\rm H\alpha}$(t) stellar time-series, which defines a global correlation that is sensitive to both short-term and long-term variability. The uncertainty on this global correlation was computed as follows: we produced 1000 realisations of synthetic time-series in which each value was replaced by the original value plus a noise derived from the uncertainty at each point (using a normal distribution). The standard deviation computed over the 1000 synthetic correlations provides the 1-$\sigma$ uncertainty on our global correlation. 

Figure~\ref{exemple} shows three typical examples of time-series and correlations  between  S$_{\rm Ca}$ and  S$_{\rm H\alpha}$. HD215152 (upper panel) exhibits a behaviour similar that of the Sun \cite[][]{livingston07,meunier09a}, with a  good correlation between  the two indexes. The other extreme can be observed with HD7199 (lower panel), with a strong anti-correlation between the two indexes. 
The middle panel shows another typical example in our sample: in this case, the Ca II emission varies with time, with an amplitude lower than the solar one but with a very significant cycle-like behaviour, while the H$\alpha$ emission does not exhibit any cycle-like variations. There are several similar stars in our sample, showing a cycle-like variability in Ca II and  no long-term variability in H$\alpha$. This is also very puzzling and we attempt to understand whether this is consistent with a variability caused by plages below.

To more easily investigate the causes of these different behaviours, we defined three categories representative of those examples, selecting stars with a good temporal sampling and a large number of observations: correlation close to zero with significant long-term Ca II variation (category \#1, 16 stars), anti-correlation (category \#2, 13 stars), and correlation (category \#3, 20 stars). 
Figure~\ref{ca_categ} in Appendix E.1 shows that the three categories are well represented at all T$_{\rm eff}$. Stars from categories \#1 and \#2 are less active in terms of average $\log R'_{HK}$, while stars in category \#3 are more active. We analysed both the whole sample and these three individual categories  in more detail. Let us note that many stars are not in these categories either because they have a poorer sampling or because they have an intermediate behaviour (e.g. moderate correlation).

\subsubsection{Long-term and short-term definitions}

In addition to the global correlation, we investigated the variability at two different temporal scales, short-term and long-term. Different conditions and thresholds have been tested and they show consistent results.
The short-term (ST) variability of the indexes is expected to be  caused by the rotational modulation of the contribution of plages. To study this timescale, we identified subsets extracted from each time-series whenever possible, that is those with enough points and a short temporal coverage, hereafter denoted zooms or subsets: these are extracts of one time-series with at least five observations not separated by more  than 20 nights, and up to a maximum time of 100 nights, which is necessary  to avoid the effect of long-term trends (this threshold allows us to cover several rotations for those stars, while maintaining a good compromise with the number of points). For the stars in our sample, this upper limit typically covers a few rotations: a conversion of the average $\log R'_{HK}$ and B-V values into a rotation period according to \cite{noyes84} shows that for most stars for which we study such ST variability, the period is comprised of between 10 and 50 days, and is shorter for 11\% of them. 
The zooms are chosen not to overlap in order to ensure that they are independent in the subsequent analysis. Due to this upper limit, the temporal sampling impacts a few zooms, but in practice most of them are not impacted by this limit: the median of the spans is 29 days, and  only 7\% of them comprise between 90 and 100 days.  
This leads to a total of 1145 zooms for 320 stars. Of these stars, 68 have at least five zooms, and 30 stars have more than ten. The median temporal coverage of the zooms is 29 nights.
The median number of nights per zoom is 9, with values of between 6 and  70.  
For each zoom, we computed the correlation as previously, which allows us to compare the short-timescale variability of Ca II and H$\alpha$. In the following, we analyse both these individual values, as well as the average of all ST correlations for each star. The results are analysed in Sect.~4.4. The same zooms were used to characterise the amplitude of the ST variability in Sect.~4.5. 

Finally, we also considered the long-term (LT) variability of these stars. This variability sometimes takes the shape of a solar-like cycle, as illustrated on Fig.~\ref{exemple}, but fitting this variability with a single sinusoid as done in \cite{gomes14} does not seem adequate here for two reasons. First, some time-series can cover two cycles with different amplitudes, which cannot be properly fitted by  a single sinusoidal function. Second, some  stars in our sample show significant LT variability but no cyclic behaviour. Fitting more sinusoidal functions at different periods would be possible for a few stars with very good temporal coverage, but would not be relevant for most of these stars because it would involve too many parameters. We  therefore chose to identify subsets of the time-series, hereafter referred to as seasons, defined as periods separated by more than 60 nights (the  gap is longer than for the ST definition to guarantee that seasons are sufficiently well separated): the average properties over each season is then used to study the LT  variability. These seasons include between 1 and 120 nights. Of our sample, 402 stars have at least four seasons. With the additional constraint that seasons must include at least five nights, we obtained 152 stars with at least four seasons, which is our basis for the analysis of the  LT  variability. 
For this selection, the median time-span of each season is 95 nights, with values varying between 5 and 316 nights. The median number of nights per season is 10. For each season, we computed the average of the indexes and their uncertainty, and then the correlation between Ca II and H$\alpha$ as previously, leading to our  LT  correlation.  
Uncertainties were computed as for the global correlation. The results are analysed in Sect.~4.4. The same seasons are used to characterise the amplitude of the LT  variability in Sect.~4.5. The values of the correlations are shown in Table~\ref{tab_sample}.

\begin{table}
\caption{Typical S/N in Ca II and H$\alpha$}
\label{tab_sn}
\begin{center}
\renewcommand{\footnoterule}{}  
\begin{tabular}{llllll}
\hline
S/N & Cat. 1 & Cat. 2 & Cat. 3 & All & Sel. Fig. {\rm 10} \\ \hline
Ca global & 7.3 & 7.5 & 12.5 & 3.7 & 11.8 \\
Ca LT & 7.9 & 7.8 & 10.4 & 4.5 & 8.0 \\
Ca ST & 3.2 & 2.2 & 5.11 & 2.3 & 4.2 \\
H$\alpha$ global & 2.7 & 3.0 & 4.7 & 3.0 & 3.7 \\
H$\alpha$ LT & 2.4 & 3.0 & 4.2 & 2.3 & 3.5 \\
H$\alpha$ ST & 2.1 & 1.7 & 2.8 & 1.9 & 2.3 \\
\hline
\end{tabular}
\end{center}
\tablefoot{S/N computed for different categories of stars (see Sect.~4.1.1), for the two chromospheric indexes at different timescales. The selection from Fig.~\ref{fig_ratio_obs} (last column) corresponds to the selection of the most variable stars in the middle panel (i.e. LT  Ca rms higher than 0.01). 
}
\end{table}

The presence of noise generally affects the estimation of the correlations and amplitudes shown in the following sections. 
The noise is therefore taken into account in the models in Sect.~5. Table~\ref{tab_sn} summarises the S/N for different categories of stars and at different timescales, for both Ca II and H$_\alpha$. The S/N is different for the various categories (higher for category \#3, which exhibits a higher variability compared to \#1 and \#2), for the two emission indexes (higher for Ca II compared to H$\alpha$), and for different temporal scales (higher for global and LT  analysis compared to ST analysis). When selecting time-series with a high LT amplitude in Ca, which is also characterised in Sect.~4.5, S/N values are intermediate between  categories \#2 and \#3, with similar results between ST  and LT  analyses, and between Ca II and H$\alpha$. 


\subsection{Significance of the temporal variability}

\begin{table}
\caption{Statistics on variability}
\label{tab_var}
\begin{center}
\renewcommand{\footnoterule}{}  
\begin{tabular}{lllll}
\hline
p$\chi^2$  & Ca & H$\alpha$ & Combined & At least one  \\
\hline
\multicolumn{5}{c}{Global (441 stars)} \\ \hline
1\%  & 434 (98 \%) & 429 (97 \%) &  422 (96 \%)& 441 (100 \%)\\
5\% &  437 (99 \%) & 435 (99 \%) & 413  ( \%) & 441 (100 \%)\\
\hline
\multicolumn{5}{c}{LT (152 stars)} \\ \hline
1\%  & 143 (94 \%) &  123 (81 \%) & 118 (78 \%) & 148 (97 \%)\\
5\% & 146 (96 \%) & 132 (87 \%) &  128 (84 \%)& 150 (99 \%)\\
\hline\end{tabular}
\end{center}
\tablefoot{Number of stars with $\chi^2$ probability (p$\chi^2$) lower than the indicated threshold, indicating significant variability, for various timescales: all timescales combined (global, based on all-night time-series), and LT, based on stars with at least four seasons, each of at least five nights. Percentages are indicated in brackets (based on 441 and 152 stars respectively). The "Combined" column corresponds to low probabilities in both Ca and H$\alpha$.  
}
\end{table}

For each time-series, the reduced $\chi^2$ is computed, which confers an advantage over the F-test probability in that it takes  into account the heterogeneous uncertainties over a time-series. The $\chi^2$ probability (p$\chi^2$) is deduced from the $\chi^2$ and the number of points. A low value of p$\chi^2$ indicates significant variability given the estimated uncertainties (i.e. the constant model is not sufficient to explain the observations). The statistics for our sample of 441 stars is shown in the first half of Table~\ref{tab_var}, separately for both indexes, and then combined (in that case we impose that both probabilities should be below the threshold) or in at least one of the indexes. We conclude that most stars show significant variability (98\% in Ca II and 97\% in H$\alpha$, at the 1\% level) at least on some  timescale. 

We applied the same analysis to all seasons, keeping only stars with more than five seasons with at least four nights. At the 1\% level, 94\% and 81\% of the stars show significant LT variability in Ca II and H$\alpha,$ respectively, 78\% when combined. Almost all stars exhibit significant variability in at least one of the indexes. There are more stars that show significant Ca II variability than stars showing significant variability in H$\alpha$, which may be due to a larger S/N for Ca II (see Table~\ref{tab_sn}): there are 14 stars (9 \%) showing significant variability in Ca II ($<$1\%) associated to a H$\alpha$ probability of higher than 10\%, while there are  only 2 stars showing a significant variability in H$\alpha$ ($<$1\%) associated to a Ca II probability higher than 10\%. Given that there is more flux in the H$\alpha$ line compared to the Ca II lines, this means that the chromospheric emission is intrinsically larger in Ca II compared to H$\alpha$ for F-G-K stars.

We conclude that a large fraction of stars in our sample exhibit significant variability. Furthermore, some stars show significant variability in Ca II and not in H$\alpha$, and vice versa, which suggests a complex relationship between the two types of emission; this is studied in more detail in the following sections.


\subsection{Observed global correlations from nightly time-series}

\begin{figure}
\includegraphics{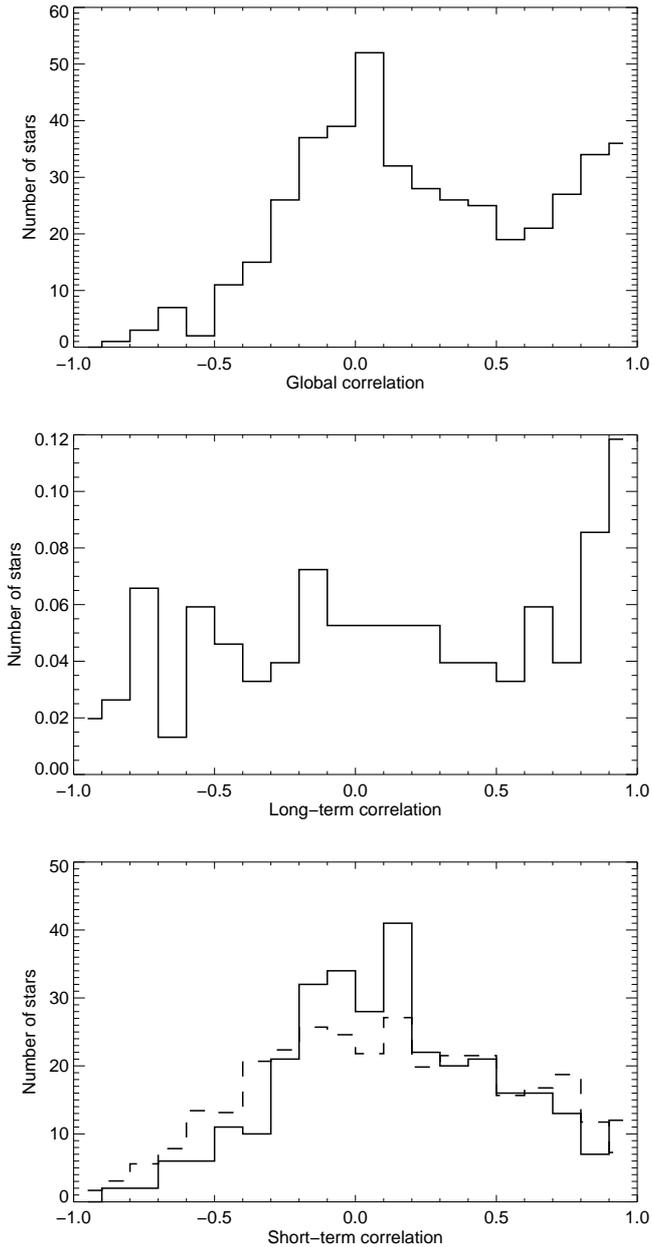}
\caption{
Distribution of the correlation values at different temporal scales: global correlation (upper panel), LT  correlation (middle panel), individual ST correlations in solid line, and average (for each star) ST correlations  in dashed line (lower panel). 
}
\label{distcorrel}
\end{figure}

\begin{figure}
\includegraphics{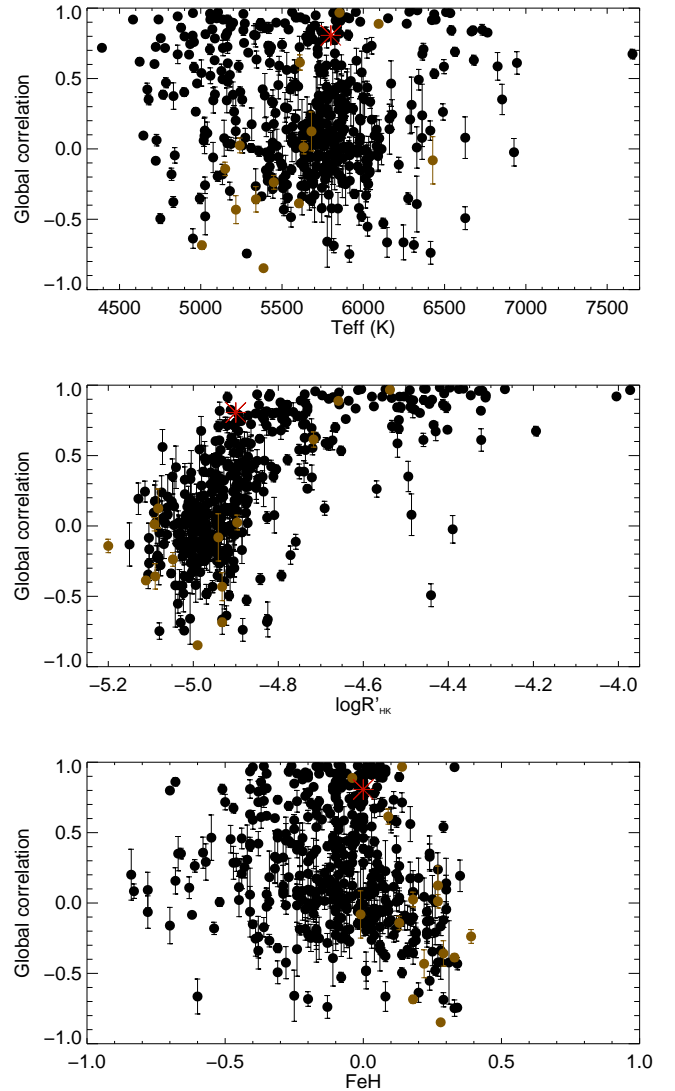}
\caption{
Global correlation versus stellar parameters.  {\it Upper panel:} Global correlation for the 441 stars versus T$_{\rm eff}$. Subgiants are shown in brown. The red star indicates the  position of the Sun \cite[correlation from][]{meunier09a}.   
{\it Middle  panel:} Same versus  $\log R'_{HK}$.
{\it Lower panel:} Same versus  metallicity. Brown dots are the subgiants.
}
\label{correl}
\end{figure}

We first analysed the global correlation computed for each of the 441 stars, the distribution of which is shown as a solid black line in Figure~\ref{distcorrel}. We observe a peak close to a correlation of 1 (stars in the solar regime, correlations typically larger than 0.6), and another wide peak around zero, including stars with a low uncertainty on the correlation: 24 stars have an absolute value of the correlation lower than 0.2 at the 3-$\sigma$ level.
Of the 441 stars, 137  have a correlation higher than 0.5 (31\%, the Sun would be in this category), and 192 (44\%) have a positive correlation different from 0 at the 3-$\sigma$ level. The tail of the distribution extends to anti-correlated stars, with 13 stars (3\%) having a correlation below -0.5. Also, 54 stars (12\%) have a negative correlation, different from 0 at the 3-$\sigma$ level. These results are compatible with those of \cite{gomes14}. 
Figure~\ref{correl} shows the global correlation as a function of stellar parameters and average activity level. We do not observe any significant trend with temperature, suggesting that the spectral type is not an important factor. On the other hand, we observe a strong correlation with the average activity level, as the correlations close to zero are found mostly for low-activity stars, as are the anti-correlations. There are almost no  active stars with a well-defined anti-correlation or a correlation close to zero; although we have far fewer stars in this activity regime.
Finally, there is a very weak anti-correlation between the global correlation and the metallicity. However, this result is difficult to interpret because of the strong biases in our sample, and the fact that stars with a low  $\log R'_{HK}$ tend to have a higher metallicity as well.

\subsection{Observed long-term and short-term correlations}

\begin{table}
\caption{Statistics on correlations at different timescales}
\label{tab_correl}
\begin{center}
\renewcommand{\footnoterule}{}  
\begin{tabular}{llll}
\hline
Correlation  & Global & LT & ST \\
sign & (441 stars) & (152 stars) & (320 stars)\\ 
\hline
Positive & 192 (44\%) & 86 (57\%) & 61 (19\%) \\
Negative & 54 (12\%) & 39 (26\%) & 9 (3\%) \\
\hline
\end{tabular}
\end{center}
\tablefoot{Number of stars and percentage with a positive or negative correlation at the 3-$\sigma$ level for the different time-scales. 
}
\end{table}

\begin{figure}
\includegraphics{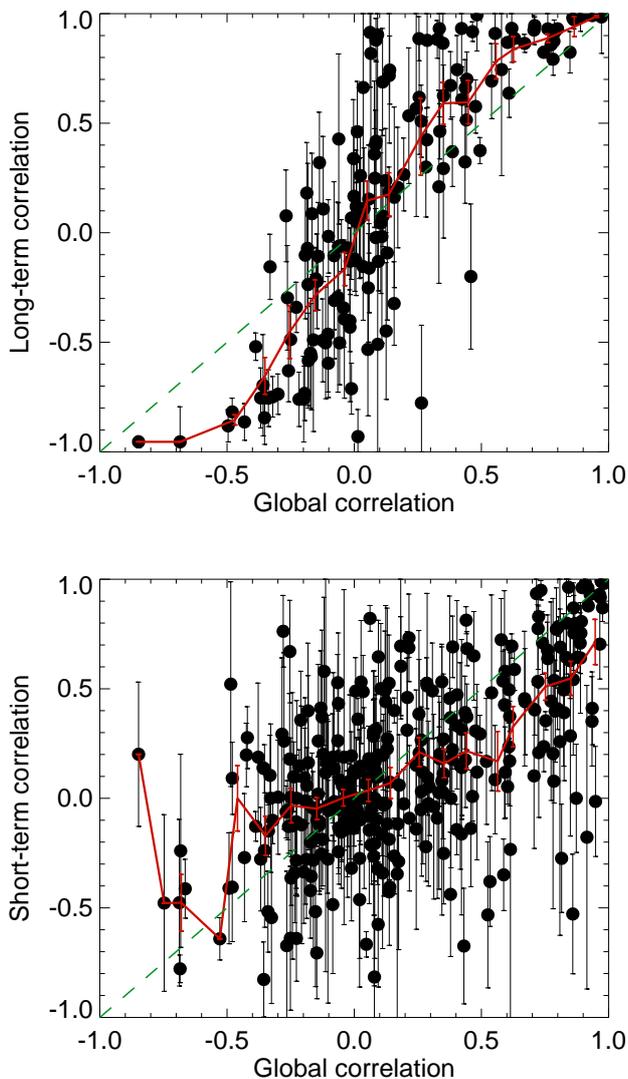}
\caption{
{\it Upper panel}: LT  correlation versus global correlation (black). Bins in global correlation of 0.1 in width are defined, and LT correlations are averaged over each bin (red line).  The green dashed line is the y=x line.  
{\it Lower panel}: Same for the ST correlation. 
}
\label{correlall}
\end{figure}

We computed the LT  correlations and  ST correlations for subsamples of stars. First, the LT  correlations were computed for a subsample of 152 stars respecting the conditions described in Sect.~4.1.2, that is, with at least four seasons of more than five nights. The number and percentage of stars with a positive or negative correlation are shown in Table~\ref{tab_correl} and compared to the rates for the global correlation.
The LT  correlations are shown as a function of the global correlation in the upper panel of Fig.~\ref{correlall}. 
We then considered bins in global correlation of  0.1 in width and averaged the LT  correlation of all stars in each bin: the result is shown by the red curve.
There is a trend for the strong correlation and anti-correlation to be reinforced when considering the LT  behaviour, at least for intermediate correlations; 
when very close to 1 (or -1), the two are very similar, and there is
little margin to get closer to 1 (or -1).
This is also observed with the distribution of correlations in Fig.~\ref{distcorrel}: a Kolmogorov-Smirnov test between the global correlation and this LT  correlation indicates a very small probability (0.0007) that the two distributions originate from the same parent distribution.

Concerning the ST correlations, the number and percentage of stars with a positive or negative correlation are also  shown in Table~\ref{tab_correl} and are compared to the rates for the global correlation.
These ST correlations are shown as a function of the global correlation in the lower panel of Fig.~\ref{correlall} for the subsample of 320 stars, where each point represents the average over the zooms of a star). 
As for the LT  correlations, the ST correlations are then averaged in bins in global correlation (red curve):  we observe a trend for the ST correlations to be closer to zero compared to the global correlation. In particular, stars with a strong global anti-correlation tend to have a lower degree of ST anti-correlation, suggesting that the causes of the anti-correlation mostly  affect the chromospheric indexes on long timescales.  The distributions are compared in Fig.~\ref{distcorrel}. The Kolmogorov-Smirnov test also shows that the probability that the two distributions come from the same parent distribution is very small: 3 10$^{-5}$ for the individual correlations. However, this value is 0.001 for the average correlation. For comparison, \cite{meunier09a} shows that for the Sun, which has a global correlation close to 0.8 and is dominated by the LT  behaviour, the  solar ST correlation computed over 1-2 years only  could be close to zero or slightly negative, as observed here. 

The individual correlations in zooms (not shown here) show a large dispersion, which appears to be significant in some stars: when selected with a threshold at five zooms per star, the dispersion of the individual correlation is significant at the 1\% level (according to an F-test applied to the rms of the individual correlation of a given star and its average uncertainty) for 12\% of the stars (29\% for a threshold of 5\% probability); the same is true for 23\% of the stars when selecting stars with more than ten zooms. For some stars, the zoom correlations have the same sign as the global correlation, but in other cases the sign can be different for certain zooms.

We conclude that there is a global agreement between the LT  and ST correlations with the global correlation. However, the LT correlations tend to be closer to 1 or -1,  while the ST correlations are closer to zero.

\subsection{Observed H$\alpha$ variability compared to Ca II}

\begin{figure}
\includegraphics{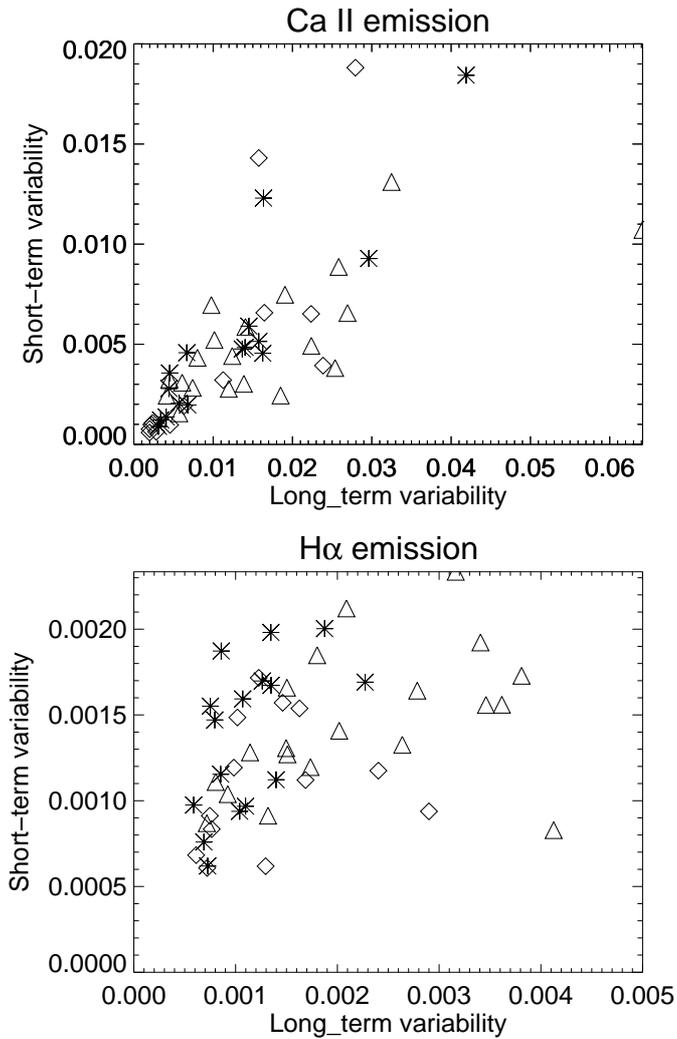}
\caption{
ST variability versus LT variability for Ca II (upper panel) and H$\alpha$ (lower panel) for stars of the three categories: \#1 (stars), \#2 (diamonds), and \#3 (triangle). 
}
\label{ca_ha}
\end{figure}

\begin{figure}
\includegraphics{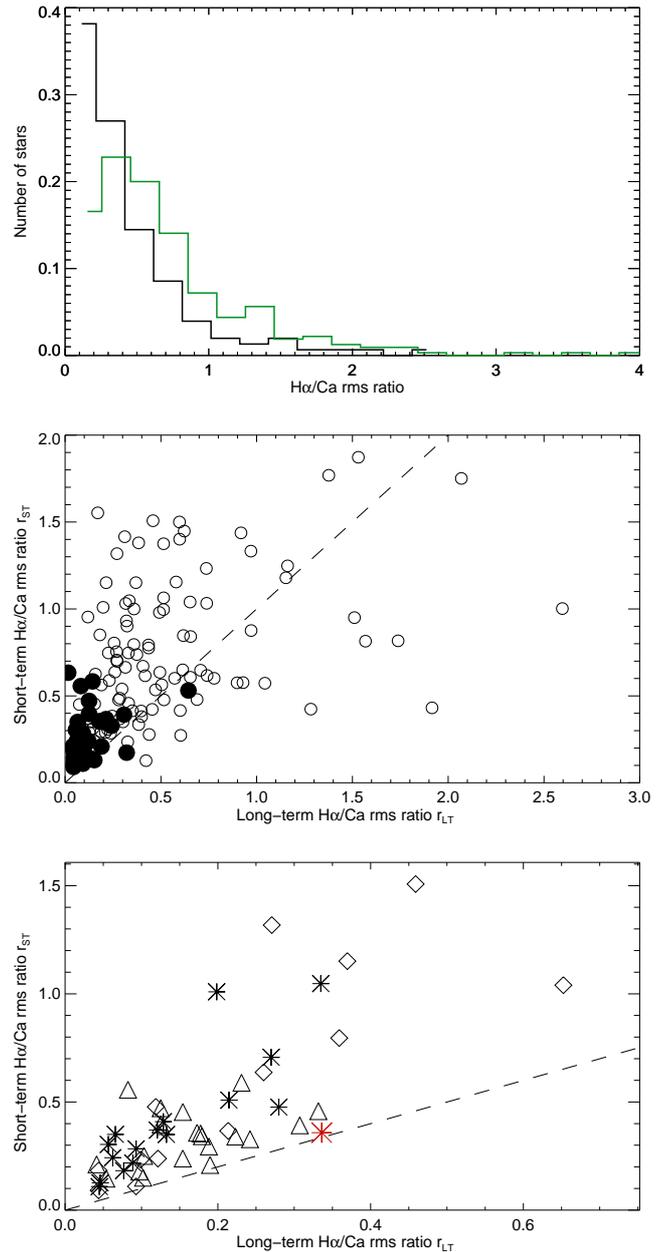}
\caption{
{\it Upper panel}: 
Distribution of ratios between H$\alpha$ and Ca II rms emissions, $r_{\rm LT}$ (black), and $r_{\rm ST}$ (green). The  last one can reach values up to 7.1, but the plot has been cut to ratios of at most 4 for clarity, as there are very few stars with a higher ratio. 
{\it Middle panel}: ST rms ratio versus LT  rms ratio (for stars common to the two subsamples), for stars with high LT  Ca amplitude (rms$>$0.01, filled circles), and stars with a lower LT variability (open circles). The dashed line is the y=x line. 
{\it Lower panel}: Same for stars in category \#1 (stars), category \#2 (diamonds), and category \#3 (triangles). The Sun is indicated by the red star.
}
\label{fig_ratio_obs}
\end{figure}

To quantitatively compare the variability of the two chromospheric indexes, we computed the ratio between the H$\alpha$ and Ca II variabilities on both timescales.  The rms on different timescales is shown in Fig.~\ref{ca_ha}: there is a reasonable correlation between the Ca II ST and LT   variabilities (0.72 over the 49 stars in the three categories). The correlation between the two is poorer in H$\alpha$ (0.39), with more dispersion in the ST variability. We also defined the LT  ratio $r_{\rm LT}$ as the ratio between the rms of S$_{\rm H\alpha}$ for seasons satisfying the same criterion as before (152 stars) and the rms of S$_{\rm Ca}$ for the same seasons. The ST ratio $r_{\rm ST}$  was defined as the rms of S$_{\rm H\alpha}$ and the rms of S$_{\rm Ca}$ over all available zooms for a given star, after subtraction of the average in each zoom (which corresponds to the rms of the residuals).

The distributions of the two ratios are shown in Fig.~\ref{fig_ratio_obs} (upper panel). The ratio covers a wide  range of values, but the large values are predominantly due to stars with a low LT  variability in Ca, and therefore a low S/N, as shown in the middle panel by the filled circles. 
On the other hand, the LT  ratio distribution seems to peak at lower values than the ST ratio. 
This larger ST ratio appears significant. It is observed both for stars with a strong LT  variability (middle panel), but also for the selection of stars with a very good temporal sampling and different properties in terms of correlation (categories \#1 no correlation, \#2 anti-correlation, and \#3 correlation), as shown in the lower panel: all points lie above the y=x line, including those for categories \#2 and \#3 for which a LT  variability is also present in H$\alpha$. The median of $\frac{r_{ST}}{r_{LT}} $
is 3.8 (category \#1), 3.1 (category \#2), and 2.6 (category \#3). In the following section, we present simulations that are designed to determine whether such differences can be explained by plage properties alone, as we would expect those to produce similar ratios for ST and LT  variability. 
 
For comparison, we computed $r_{ST}$ and $r_{LT}$ with similar criteria for the Sun, using the Ca II and H$\alpha$ series obtained at Kitt Peak \cite[][]{livingston07}, and whose correlations were studied  in \cite{meunier09a}. We defined 39 subsets of 100 days over a solar cycle, each including at least five points. The resulting $r_{LT}$  is 0.336, and $r_{ST}$ is 0.358 (indicated by a red star in the lower panel in Fig.\ref{fig_ratio_obs}): The ST ratio is therefore slightly higher for the Sun as well, but much less than in the observations of this large stellar sample.


\section{Can plages explain the relationship between Ca II and H$\alpha$ emissions ?}

A detailed analysis of the relationship between Ca II and H$\alpha$ emissions for a large sample of stars has not only shown that they could be anti-correlated (compared to correlations in the solar case), but also that the correlation  depends on the timescale: for example, ST correlations tend to be closer to zero than LT  correlations. 
When measured relatively to Ca II, the H$\alpha$ variability appears to be larger on short timescales than on longer timescales. Furthermore, we observed that some stars have a strong LT  Ca II variability (similar to the Sun) but no detectable variability
in H$\alpha$: could this be due to a strong ST plage signal in H$\alpha$ (masking any LT  variability), or, conversely, to a weaker  H$\alpha$ emission in plages in such stars?

To investigate this, we assumed that the H$\alpha$ variability is solely due to plages and explored three hypotheses for the Ca II--H$\alpha$ emission in plages: 
\begin{itemize}
    \item Hypothesis A (described in detail in Appendix ~C.1) assumes that all plages for all stars in a given T$_{\rm eff}$ range behave similarly, following stars with  an excellent correlation between Ca II and H$\alpha$ (as in the Sun).
    \item Hypothesis B (described in Appendix~C.2) assumes that the Ca II--H$\alpha$ relationship depends on each star and in particular on its activity level, according to laws derived from the behaviour of the complex average flux--flux relation (Sect.~3).
    \item Finally, Hypothesis C (described in Appendix~C.3) assumes that the  H$\alpha$ / Ca II plage emission ratio is constant for a given star and is given by the LT  slope between the two indexes for that star, while it differs from one star to another.
\end{itemize}  
Such laws were used to derive expected H$\alpha$ time-series, either from the observed (Appendix C.4) or synthetic (Appendix C.5) time-series representative of the stars in our sample. We now compare the correlations and amplitudes obtained with the modelled H$\alpha$ time-series for these three hypotheses.

\subsection{Correlations obtained with synthetic H$\alpha$ derived from observed S$_{\rm Ca}$}

\begin{figure}
\includegraphics{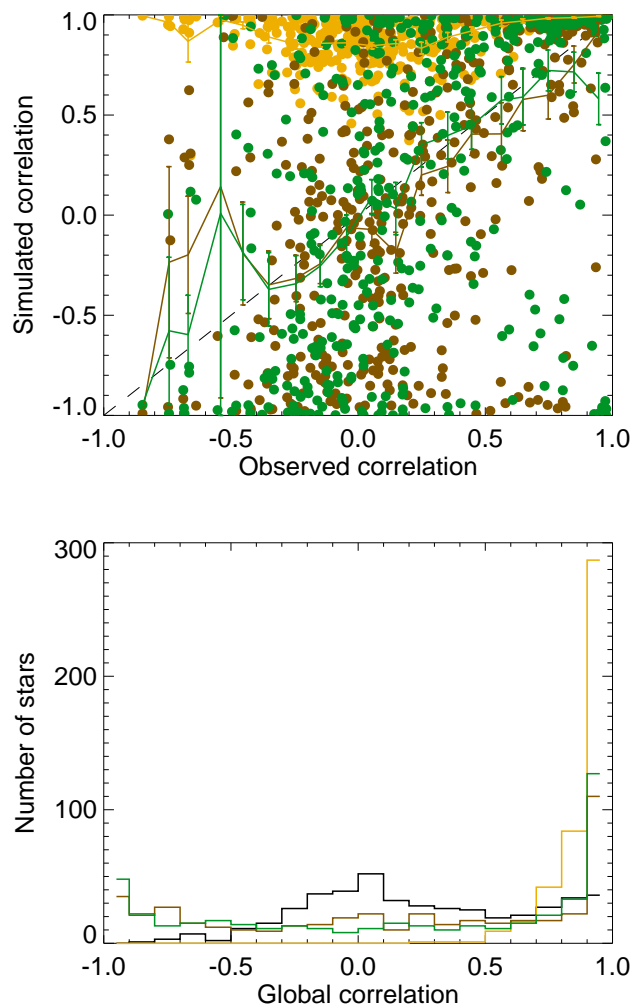}
\caption{
{\it Upper panel:} Simulated global correlation (H$\alpha$ reconstructed from observed Ca II) for all stars versus observed correlation, for hypothesis A (orange), hypothesis B (brown), and hypothesis C (green). 
{\it Lower panel:} Distribution of global correlation: observed (black), hypothesis A (orange), hypothesis B (brown), and hypothesis C (green). 
}
\label{correl_synthobs_all}
\end{figure}

\begin{figure}
\includegraphics{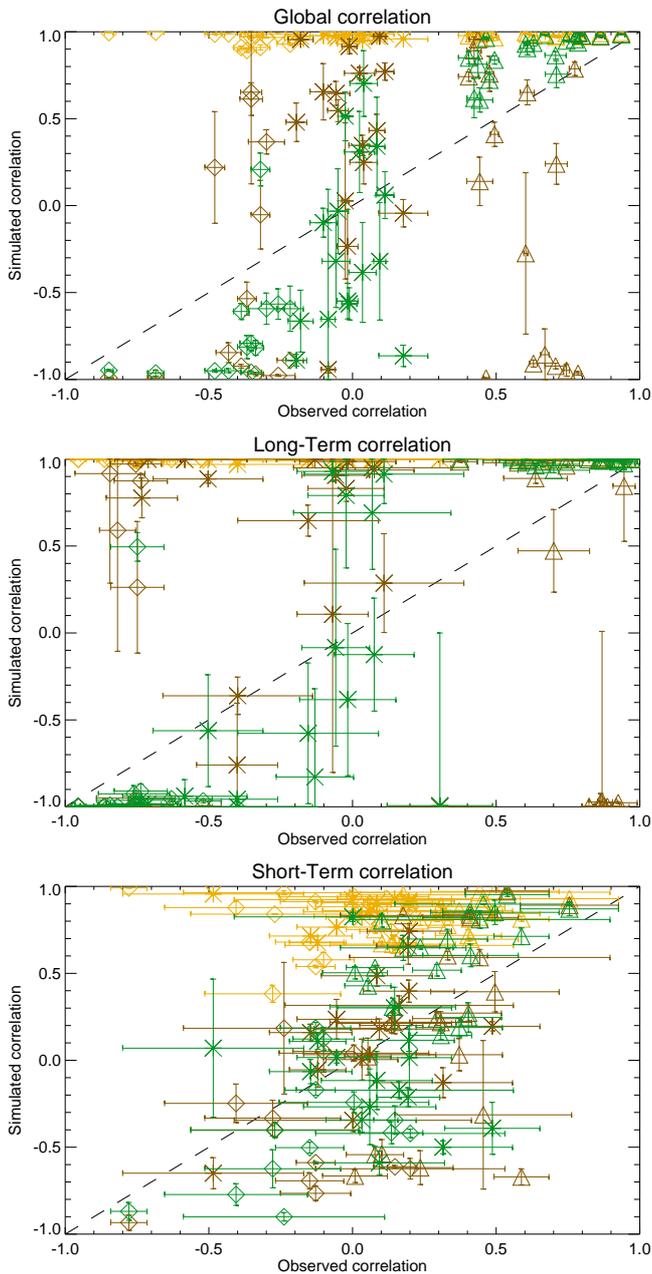}
\caption{
Simulated correlation (H$\alpha$ reconstructed from observed Ca II) for all stars versus observed correlation, for hypothesis A (orange) and B (brown) for different definitions: global correlation (upper panel), LT  correlation (middle panel), and ST correlation (lower panel). Different symbols correspond to different categories: category \#1 (stars), category \#2 (diamonds), and category \#3 (triangle). The errorbar-like symbols in the y direction represent the range covered when taking the uncertainties on the H$\alpha$-Ca II laws into account (see Appendix ~C.1 to C.3).
}
\label{correl_synthobs_categ}
\end{figure}

We first reconstructed H$\alpha$ synthetic time-series using observed S$_{\rm Ca}$ time-series as described in Appendix C.3, based on  hypotheses A, B, and C for plage properties. The correlations (global, LT, and ST) were computed as in the previous section. We compared them with observed correlations, for the full sample and for selected stars in the three categories defined in Sect.~{4.1.1}.

Figure~\ref{correl_synthobs_all} shows the global correlations between observed Ca II time-series and synthetic H$\alpha$ time-series. Hypothesis A produces correlations which are almost always higher than 0.5, with no anti-correlation nor correlation close to zero, despite the fact that the synthetic time-series take  into account the poor sampling, the low number of points, and the noise. Hypothesis B can produce anti-correlations or nearly null correlations; however, it produces very few correlations close to zero and an excess of anti-correlations. Additionally, the simulated time-series globally fail to reproduce the observed correlations. Indeed, most simulated stars in category \#1 exhibit a simulated correlation close to 1 (and a few with negative correlations) instead of a zero correlation. A significant number of simulated stars in category \#2 have a  positive correlation, and simulated stars in category \#3 produce anti-correlations instead of positive correlations. Similar observations can be made for Hypothesis C, except than in this case the sign of the correlation (for values far from 0) are always correct, by construction. 

The LT  correlations have  similar properties, with correlations even closer to 1 or -1, including stars in category \#1 (for which the observed correlations are very close to zero). Again, the simulated correlations for several stars in categories \#2 and \#3 do not have the correct sign either for hypotheses A or B, while they are correct for hypothesis C. 
The ST correlations are too large for hypothesis A; Hypotheses B and C produce similar disagreement with observations for ST correlations to that seen for global and LT  correlations.

In summary, hypothesis A leads to a strong disagreement with observations for categories \#1 and \#2, and correlations are too close to 1 for category \#3, meaning that some source of departure from the unique law is also visible for this category of stars. 
Hypothesis B does not provide the proper sign of the correlation in several cases of categories \#2 and \#3, while there is good agreement for a few stars in category \#2. For stars in category \#1, the synthetic H$\alpha$ time-series tend to be too strongly correlated instead of having a correlation close to 0. 
Finally, hypothesis C provides the proper sign for categories \#2 and \#3, but it is difficult to reproduce a good ST correlation between observed and synthetic H$\alpha$ time-series for all stars, even for category \#3 which has the best S/N. This lack of correlation with the observed H$\alpha$ time-series is also seen for category \#1. 
 We note that the lack of correlations close to zero could in part be due to the fact that here, all plages of a given star have the same properties (which requires more complex simulations, beyond the scope of this paper), but we do not expect this to strongly affect  the amplitudes, showing that it is important to consider different complementary diagnoses to test our assumptions. \\

We also reconstructed some H$\alpha$ time-series by considering simulated instead of observed Ca II time-series, as described in Appendix C.5. The results are very similar to those shown in Fig.~\ref{correl_synthobs_all} and Fig.~\ref{correl_synthobs_categ}. In addition to confirming the results, this shows that the noise on the observed Ca II time-series (leading to some correlated noise in the H$\alpha$ synthetic time-series, although some noise is naturally added to all H$\alpha$ synthetic time-series) does not affect our conclusions. We conclude that the models based solely on plages fail to reproduce the observations. This  indicates the presence of one or several additional processes (one of which could be the presence of filaments).


\subsection{Expected H$\alpha$ variability from  S$_{\rm Ca}$}

\begin{figure}
\includegraphics{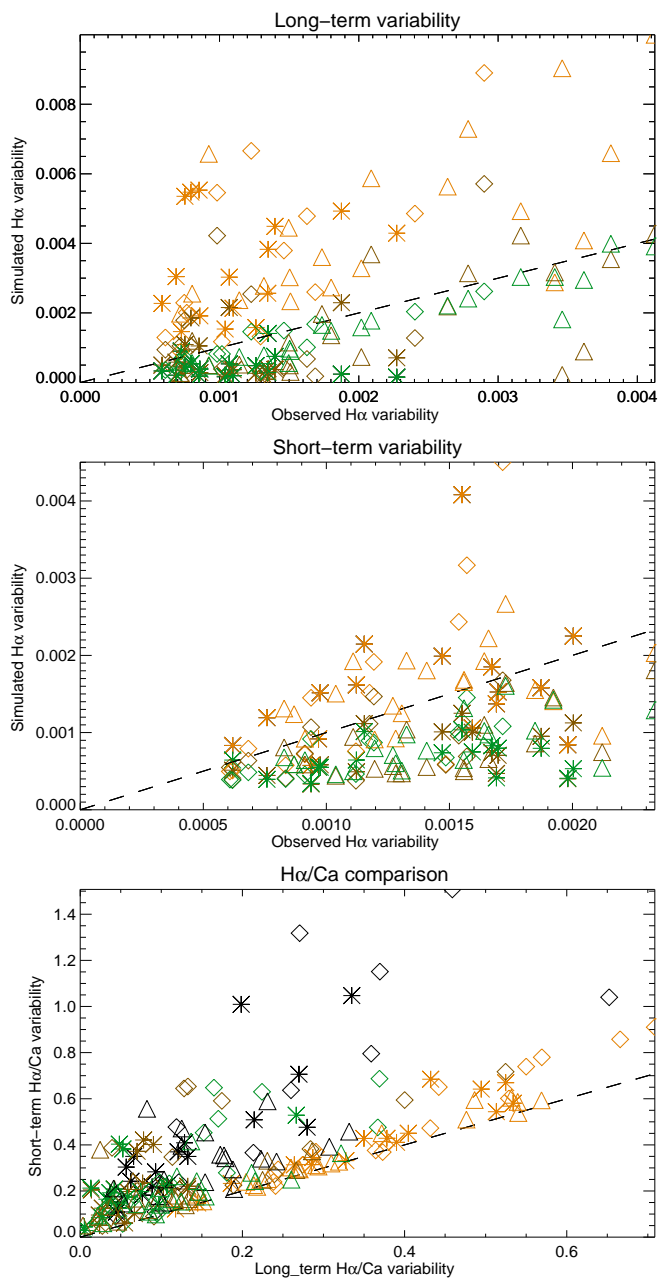}
\caption{
{\it Upper panel}: Simulated H$\alpha$ ST variability (reconstructed from observed Ca II) versus observed variability, for plage hypothesis A (orange), hypothesis B (brown), and hypothesis C (green), and for the three categories of stars: category \#1 (stars), category \#2 (diamonds), and category \#3 (triangle). The black dashed line is the y=x line.
{\it Middle panel:} Same for the LT  H$\alpha$ variability.  
{\it Lower panel:} Simulated ST H$\alpha$/Ca versus simulated LT  H$\alpha$/Ca, for the three categories of stars (same symbols and colour code). The black symbols correspond to the observed values for comparison. 
}
\label{hamod}
\end{figure}

\begin{table}
\caption{Simulated H$\alpha$ variability}
\label{tab_var_ha}
\begin{center}
\renewcommand{\footnoterule}{}  
\begin{tabular}{llllll}
\hline
Ratio & Cat. 1 & Cat. 2 & Cat. 3 & All Cat. & All \\ \hline
Synth/Obs LT (A) & 2.83 & 2.59 & 2.11 & 2.33 & 1.54\\
Synth/Obs LT (B) & 0.36 & 0.33 & 0.77 & 0.54 & 0.50 \\
Synth/Obs LT (C) & 0.45  &  0.83 &  0.82 & 0.67 & 0.63 \\
Synth/Obs ST (A) & 1.12 & 1.16 & 1.08 & 1.11 & 1.01 \\
Synth/Obs ST (B) & 0.57 & 0.80 & 0.16 & 0.59 & 0.68 \\
Synth/Obs ST (C) & 0.50  &  0.65 & 0.59  & 0.56 & 0.70 \\
\hline
H$\alpha$/Ca (Obs) LT & 0.12& 0.21& 0.17 & 0.13 & 0.32\\
H$\alpha$/Ca (A) LT & 0.38 & 0.43& 0.29 & 0.35 & 0.48\\
H$\alpha$/Ca (B) LT & 0.07& 0.13& 0.12 & 0.09 & 0.18 \\
H$\alpha$/Ca (C) LT & 0.05 & 0.15 & 0.12  & 0.10 & 0.20 \\
H$\alpha$/Ca (Obs) ST & 0.35 & 0.48 & 0.34 & 0.35 & 0.57\\
H$\alpha$/Ca (A) ST & 0.63 & 0.47 & 0.31 & 0.37 & 0.59\\
H$\alpha$/Ca (B) ST & 0.21 & 0.38 & 0.17 & 0.21 & 0.40 \\
H$\alpha$/Ca (C) ST & 0.19 & 0.28 & 0.18  & 0.20 & 0.43\\
\hline
\end{tabular}
\end{center}
\tablefoot{Comparison between synthetic and observed H$\alpha$ amplitudes (first part) and comparison of the observed and synthetic H$\alpha$/Ca ratio, for the different hypothesis for plage emission in H$\alpha$ (A, B, or C). The values indicate the median of each ratio over all considered stars. Categories are defined in Sect.~{4.1.1}.  When all stars are concerned (last column), they correspond to 152 stars for LT values and 320 stars for ST values.
}
\end{table}

We now focus on the amplitude of the reconstructed H$\alpha$ variability. We first compare the amplitude of the simulated H$\alpha$ signal directly with the observed signal on different timescales; see the first two panels of Fig.~\ref{hamod}. The analysis of the 152 stars for which the LT  amplitude can be estimated shows that hypothesis A overestimates the LT H$\alpha$ amplitude (median of the ratio between simulated and observed is 1.54), while hypotheses B and C underestimate it (median of 0.50 and 0.63 respectively). 
This is confirmed by the analysis of well-sampled stars in specific categories (median shown in Table~\ref{tab_var_ha}). The only exception is the good agreement for category \#3 (well correlated stars) for which hypothesis C gives a good agreement between observations and simulations, which is expected given that the model is based on the LT  relationship. We note that despite this assumption, the amplitudes remain underestimated for the other categories, and in particular for \#1, because there is some significant departure from a constant LT  variability in H$\alpha$  (related to the correlation being close to zero) even though the slope is close to zero. The departure from the observations is the greatest for category \#1, and the smallest for category \#3, but is present in all cases.

The ST variability was analysed in a similar way for the 320 stars for which the computation can be done. As before, hypothesis B leads to an underestimation of the H$\alpha$ amplitude (median of the ratio between simulation and observation of 0.68), as does hypothesis C (median of 0.70): the same behaviour is observed when analysing  the specific categories. On the other hand,  hypothesis A leads to a slight overestimation, but on average there is good agreement. The ratios between simulated and observed H$\alpha$ variability are much closer to 1 than for the  LT variability. Therefore, not only do the simulations based on plages fail to reproduce the observed H$\alpha$ variability, but they do so to a different extent  for the  LT  and ST variability, which does not favour the plage-only explanation. \\

Finally, we analysed the ratios between the variability in H$\alpha$ and in Ca II as done for observations in Sect.~4.5. The results are shown in the lower panel of Fig.~\ref{hamod}. The ST ratios are always higher than the LT ratios, as seen in the analysis of the observed time-series, but the factor is much smaller than what is observed in Fig.~\ref{fig_ratio_obs} for hypothesis A, that is, the values remain very close to the y=x line. Hypotheses B and C on the other hand provide a larger dispersion of ratios, but this dispersion is still smaller than in the observation.
The medians of the ratios are summarised in Table~\ref{tab_var_ha} for the three categories: the ST ratios for hypothesis A are quite similar to the observed one, but they are much smaller for hypothesis B. On the other hand, the LT ratios for hypothesis A are very high compared to the observed ones, while those for hypothesis B are more similar. None of the assumptions agree with the observations for the H$\alpha-$Ca relationship on both  short and
long timescales. 
We followed the same approach with the synthetic H$\alpha$ time-series reconstructed from realistic synthetic Ca II time-series, and observed  very similar results.

We conclude that the H$\alpha$ variability cannot be reproduced by hypothesis A (too large) or hypothesis B or C (too small). 
The synthetic H$\alpha$ time-series shows a trend for a larger  ST H$\alpha$/Ca compared to the LT  ratio as observed, but with a much smaller difference between the two temporal scales. This suggests that plages are not sufficient, because they should lead to a similar behaviour for the  LT  and ST variabilities once the uncertainties are taken into account. The fact that there is a different S/N for different categories of stars or indexes (see Sect.~4.1.2)  is nevertheless taken into account in our reconstructed time-series. The presence of noise can explain some of the observations (e.g. part of the larger H$\alpha$/Ca ST ratios compared to the LT ratios) but not for all stars. Furthermore, even if there is some uncertainty on the slopes used to build the synthetic time-series, an important conclusion here is that in many cases, no slope allows  both the LT  and ST variability to be reproduced.  This leads us to study the effect of the presence of filaments in the following section. Let us note that flares could also explain part of the observed disagreement: some possibilities are discussed in Appendix E. Nevertheless, the strong ST H$\alpha$ variability observed in several cases, which is often consistent with rotational modulation, suggests that flares cannot be the only driver of the observed behaviour.


\section{Effect of filaments on the relationship between Ca II and H$\alpha$ emissions}

We show in the previous section that H$\alpha$ time-series reconstructed from the Ca II time-series based on three different assumptions for plage properties (i.e. the H$\alpha$--Ca II relation in plages) fail to reproduce the observations in terms of correlation and H$\alpha$ variability, both on short and long temporal scales. \cite{meunier09a} proposed that the presence of filaments, which significantly modify the H$\alpha$ emission by adding an absorption contribution when filaments  are between the disk and the observer, could explain the observed anti-correlations between H$\alpha$ and Ca II emissions in a few stars (our category 2). Such filaments could also explain the presence of many stars with a correlation close to zero, even when Ca II exhibits a clear LT  solar-like variability. Here, we further explore this assumption. We first review filament properties that could be relevant to S$_{\rm H\alpha}$ emission. We then present two simple models, one focusing on  LT  variability, and the other on ST variability, as a first step in identifying parameters with a critical effect.

\subsection{Filament properties}

Our knowledge of filaments is almost exclusively based on the Sun. Their presence and properties are not known for other solar-type stars. Furthermore, important properties of stellar activity such as the spatio-temporal distribution of the activity pattern are also very poorly constrained for stars other than the Sun. This limits the interpretation of chromospheric emission properties. It is nevertheless possible to identify important 
 properties of filaments which could affect their contribution to S$_{\rm H\alpha}$ in other stars: 
\begin{itemize}
\item{{\it Filament intrinsic properties: size, contrast, number, and lifetime.} All have a direct effect on the apparent filling factor (ff) of filaments. The lifetime (especially relative to the rotation period of the star) may also affect the ST variability. There is a strong degeneracy between the four properties, as they are related to each other. For example, large filaments are more contrasted and live longer. Filament properties also vary during the solar cycle and as a function of latitude (see next point). On the Sun, the largest filaments are  seen at high latitude and last longer, sometimes more than one rotation \cite[e.g.][]{duchlev96}. We note that as there are many small filaments compared to the largest ones, their contribution to the  emission is also important: an analysis of the Meudon BASS2000\footnote{http://bass2000.obspm.fr/}  solar database shows that the apparent filling factor for filaments smaller than 20${^\circ}^2$ 
accounts for half the total filling factor. As their lifetime is short compared to the rotation period, their effect on ST correlation should be important. To our knowledge, there is little information available on stellar filaments, apart from for very young stars such as pre-MS stars \cite[e.g.][]{jardine19,jardine20}.} 
\item{{\it Filament position in latitude.}  As for spots and plages, filaments follow the butterfly diagram, with an equatorward branch similar to spots, and a poleward branch \cite[e.g.][]{ulrich02,li10,gao12}, as shown in Fig.~\ref{butterfly}. As with stellar inclination, we expect their latitude to affect the apparent filling factor and therefore potentially the correlations studied here: when seen close to pole-on, the contribution of high-latitude (large) filaments should be higher compared to the solar configuration, especially when compared to the plage contribution that occurs at lower latitudes. The difference between the latitude of filaments and that of plages is therefore an important parameter. In addition, \cite{makarov92} showed that the maximum latitude of filaments during the minimum of activity cycles is related to the amplitude of the following sunspot cycle.  However, the butterfly diagram in other stars is very ill-defined,  and we are not aware of any stellar studies concerning filaments in that context. 
}
\item{{\it Filament position in longitude}: Filament position in longitude, especially in relation with plages producing the rotation modulation of the chromospheric emission on short timescales, should affect the time-series and potentially the correlations studied in this paper. An analysis of the Meudon BASS2000 filament database alongside plages\footnote{Plages are usually defined from Ca II images corresponding to the chromospheric emission, but the definition based on magnetic flux here relies on the well-known correlation between magnetic flux intensity and Ca emission. They were defined as structures with an area higher than 300 ppm and a magnetic field threshold of 100~G} extracted from MDI/SOHO magnetograms \cite[][]{Smdi95} as in \cite{meunier18a} shows that the distribution of the differences in longitude between filaments and the closest plage (in the same hemisphere) peaks at 0$^\circ$, but is very wide. There are also many days with filaments observed on the disk and no plage (defined as above), indicating a different position in longitude; this should affect the ST correlation because of the induced phase shift of plage and filament effects in the time-series.  
However, we do not expect a strong effect on the LT  correlation of the longitude distribution, because the average contribution is considered. 
}
\end{itemize}

Finally, it is well-known from solar synoptic maps that filaments lie along lines of polarity inversion (hereafter PIL)  \cite[e.g.][]{rust66,kuperus67,hirayama85,mouradian98,ulrich02}. \cite{meunier11} quantified this relationship based on a systematic comparison of filament positions and magnetograms. However, there are more PIL pixels during cycle minimum, and less during cycle maximum because of the presence of large unipolar areas, while there are far fewer filaments during cycle minimum. This means that it is not possible to argue in terms of the amount of PIL only, because their `efficiency' in producing filaments is also critical and highly variable during the cycle: the amount of magnetic flux is likely to be important as well. The extrapolation to quiet stars is therefore not direct.

\subsection{Long-term toy models}

\begin{figure}
\includegraphics{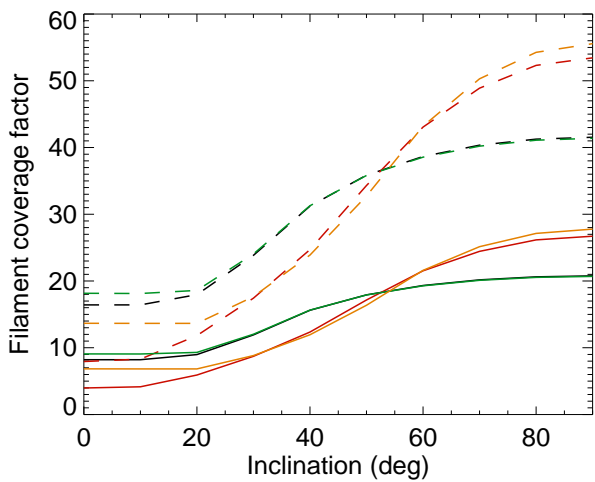}
\caption{
Factor necessary for filaments to compensate (solid lines) for the plage emission (in H$\alpha$) vs. inclination, for different configurations (Table~\ref{tab_config_lt}): LT1 (black), LT2 (red), LT3 (green), LT4 (orange). The dashed lines correspond to a filament contribution to the H$\alpha$ emission twice larger than the plage contribution (see text). 
}
\label{recap_lt}
\end{figure}

\begin{table}
\caption{LT  filament configurations}
\label{tab_config_lt}
\begin{center}
\renewcommand{\footnoterule}{}  
\begin{tabular}{lll}
\hline
Configuration & Plage latitudes & Filament latitudes   \\
 & (deg) & (deg) \\ \hline
LT1 & 16 & 36 \\
LT2 & 10 & 50  \\
LT3 & 22-9 & 40-30 \\
LT4 & 22-9 & 40-80  \\
\hline
\end{tabular}
\end{center}
\tablefoot{When two values are indicated, latitudes correspond to the beginning and end of the cycle, respectively. LT1 corresponds to average solar values. LT3 corresponds to a typical solar butterfly diagram, with filaments dominated by the equatorward branch of filament migration. LT4 considers a strong poleward branch of filament migration only. 
}

\end{table}

In this first simple model, the plage and filament coverages were modelled by latitudinal bands that are symmetric between the two hemispheres, and either fixed or migrating. These were then associated with a certain filling factor and contrast to finally compute the ratio between the contribution to H$\alpha$ emission of plages and the contribution of filaments. The model is detailed in Appendix D.1. We took the solar ratio of $\simeq$10 \cite[][]{meunier09a}  as a reference and computed the factor by which the solar filament contribution should be multiplied to compensate for the plage contribution (solid lines). 
A factor of two, for example, means that the filament contribution needs to be twice  the solar one to compensate the plages.

We present results based on the amplitude of the reconstructed emission for plages and filaments, rather than on the correlation between Ca II and H$\alpha$ emissions, because we have chosen similar variability (over the cycle) for their filling factors, which are therefore in phase and very well correlated or anti-correlated. There is therefore more information to be gleaned from an analysis of the amplitudes. 
The results are shown in Fig.~\ref{recap_lt} for a few configurations shown in Table~\ref{tab_config_lt} and versus inclination. Physically, a larger factor represents a combination of larger size, lifetime, contrast, and/or number of filaments.
A small factor means that a small departure from the solar properties is sufficient for the filaments to compensate for the plage contribution. We can see that stellar inclination has a major effect, causing a stronger relative filament contribution when the star is seen with a low inclination (and therefore a lower factor is necessary): this is because filaments are present at higher latitudes than plages. The effect is more pronounced when the difference in latitude between plages and filaments is higher. The presence of a butterfly diagram instead of a fixed latitude pattern has a smaller effect, but it may be important if the latitude drift is high. 

In order to obtain an anti-correlation, the filament contribution must  be larger than the plage contribution (and not simply compensate it); a ratio of 0.5, for example, meaning that the filament contribution is twice larger,  should be sufficient to produce an anti-correlation. Such cases are shown as dashed lines in Fig.~\ref{recap_lt}. For low inclinations, the filament contribution should be 10 to 20 times larger than the solar one to be able to produce such an anti-correlation.

In this simple model, we  took projection effects into account, but did not consider any centre-to-limb variability in plage or filament contrasts. We conclude that stellar inclination and the respective latitudes of plages and filaments are important to explain a larger contribution from filaments than from plages. A minimum contribution that is about ten times larger than solar filaments is nevertheless necessary in the most favourable configuration to be able to observe a correlation of zero or LT anti-correlations. We also note that as this can be caused by several parameters (e.g. size and contrast), a small factor for each of them could be sufficient. A factor 1.8 in size, lifetime, contrast, and number (if all are affected simultaneously), for example, would be sufficient to provide the factor ten. We also note that if the H$\alpha$ emission in a plage is lower than in the solar case (see Sect.~5), then it will also be easier for filaments to compensate for their emission, which highlights an indirect role of plages.

\subsection{Short-term toy models}

\begin{figure}
\includegraphics{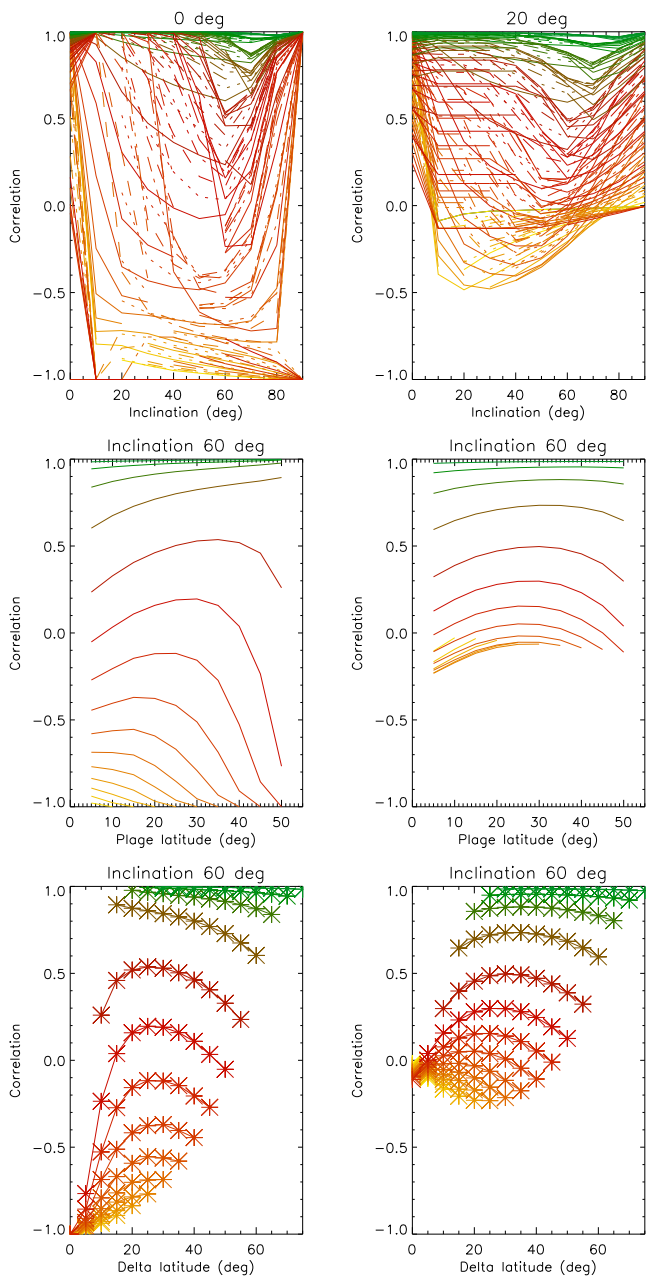}
\caption{
ST correlation versus inclination (upper panel, all latitude configurations for plages and filaments), ST correlation versus plage latitude (middle panel, inclination of 60$^\circ$ only), and difference between filament and plage latitude (lower panel, inclination of 60$^\circ$), for two longitude configurations: difference in longitude between the plage and the filament of 0$^\circ$ (left panels) and 20$^\circ$ (right panels). Colours from yellow to green correspond to increasing filament latitude in all cases. The different line styles in the upper panel correspond to different plage latitudes. 
}
\label{recap_ct}
\end{figure}

Finally, we built a very simple model based on one plage and one filament to study the effect of filament parameters on the ST correlation. This model is described in Appendix D.2. 
We note that the temporal variations in the Ca II and H$\alpha$ emission are no longer in phase. However, the correlation, which is also a variable we measure in previous sections, provides a good way to compare the time-series. We  explored the following range of parameters: plage latitudes vary between 5$^\circ$ and 55$^\circ$ (steps of 5$^\circ$); filament latitudes vary between the plage latitude and 85$^\circ$ (steps of 5$^\circ$); the difference between the plage and the filament longitudes varies between 0 and 140$^\circ$ (steps of 20$^\circ$); the ratio between the plage and the filament size takes the values 10 (solar-like), 5, 1, and 0.5. 

We find that negative correlations between the synthetic Ca~II and H$\alpha$ time-series can be obtained only for a large filament contribution (typically, 20 times more than in the solar case if plage properties remain the same), and small difference in plage and filament longitudes (0 and 20$^\circ$ cases).
We illustrate those specific cases in Fig.~\ref{recap_ct}. 

The upper panels show the dependence on inclination. The behaviour is  different from the LT properties discussed in Sect.~6.2. For 0$^\circ$ and 90$^\circ$, the correlations are either -1 or 1. In between, there is a variability with inclination, but without a single trend and in some cases a low dependence. The following panels focus on an inclination of 60$^\circ$ in order to highlight the dependence on latitude. Anti-correlations are obtained mostly for low-latitude plages,  and only when the difference in latitude between the plage and the filament is small, which is very different from the LT model. 

Even though this is a very simple model, it shows that the parameters do not have a similar effect on the LT and ST relationship. In the ST case, for example, it is better to have a small difference in latitude between plage and filament to produce an anti-correlation, while it was easier with a large difference in the LT model. This type of difference could be responsible for the difference in behaviour between LT and ST observed in Sect.~4 and 5, although more sophisticated models will be needed in the future. Furthermore, the ratio between plage and filament is related here to size and contrast, as in the LT case, but the lifetime should have a complex effect  depending on whether it is short or long compared to the rotation period of the star.

\section{Conclusion}

In this study, we first  extended the analysis of the relationships between the Ca II and H$\alpha$ emission for a large sample of F-G-K stars by considering the average emission level and the short and LT  variability. We then  investigated whether the observed properties could be reproduced by a different sensitivity of H$\alpha$ and Ca II emission to plages.  Our main findings can be summarised as followed:
\begin{itemize}
    \item We confirm that very few stars  (only 3\%) have clear anti-correlated Ca II and H$\alpha$ emissions  (12\% have a negative correlation at the 3-$\sigma$ level) and we show that this is mostly caused by LT  variations. For these stars, the ST correlation is still negative but not as strongly negative as the LT  correlation. 
    \item 20\% of the stars that show strong LT Ca II variability have a  correlation between 
Ca II and H$\alpha$ emissions that is very close to zero (at all timescales), despite showing significant H$\alpha$ variability. The mechanisms causing the anti-correlation could also be responsible for such a null correlation. 
    \item The ratio between the H$\alpha$ and Ca II ST variabilities is generally significantly larger (up to a factor 3) than the ratio between the LT  variabilities of  H$\alpha$ and Ca II,  which is not the case for the Sun. This cannot be explained by the difference in noise levels between indexes. 
    \item 29\% of the low-activity stars in the 5300-6100 K range, among which  a large fraction (40\% instead of 13\% for all quiet stars in that T$_{\rm eff}$ range) have a high metallicity, have larger H$\alpha$ emission with respect to what would be expected if its relation with Ca II emission were closely correlated. This creates two `branches' at low activity in the H$\alpha$--Ca II emission plots. However, these stars do not seem to have a preferred correlation.   
    \item Even by assuming different dependencies of the H$\alpha$ emission on plages, we were not able to reproduce the observations when considering that the ( ST and LT) variability is caused by plages only; no model was able to reproduce the variability on both timescales.
    \item Very little is known about filaments on stars other than the Sun, which means there are a significant number of free parameters when considering their effect on the stellar flux. Considering an important filament filling factor (20 times the solar one) and filaments located at different latitudes from those of plages allows us to obtain anti-correlation on long timescales. Having filaments closer to plages allows us to retrieve anti-correlation on short timescales. However, more realistic modelling of filaments is required to assess their impact on H$\alpha$ emission.     
\end{itemize}

\begin{acknowledgements}

This work was supported by the "Programme National de Physique Stellaire" (PNPS) and "Programme des relations Soleil-Terre" (PNST) of CNRS/INSU co-funded by CEA and CNES.
This work was supported by the Programme National de Plan\'etologie (PNP) of CNRS/INSU, co-funded by CNES. 
The HARPS data have been retrieved from the ESO archive at http://archive.eso.org/wdb/wdb/adp/phase3\_spectral/form.
This research has made use of the SIMBAD database, operated at CDS, Strasbourg, France. This work made use of the BASS2000 (http://bass2000.obspm.fr/) filament catalogue and we thank Jean Aboudarham and Christian Rénié for their help. 
ESO program IDs corresponding to data used in this paper are: 
183.C-0972
075.C-0202
075.C-0689
077.C-0295
184.C-0815
076.D-0103
076.C-0279
078.C-0209
080.C-0712
080.C-0664
081.C-0774
082.C-0412
091.C-0936
192.C-0852
60.A-9036
188.C-0265
080.D-0347
082.C-0212
085.C-0063
086.C-0284
190.C-0027
196.C-0042
198.C-0836
086.C-0145
074.C-0364
084.C-0229
086.C-0230
088.C-0011
183.C-0437
092.C-0579
094.C-0797
093.C-0062
095.C-0040
096.C-0053
096.C-0499
097.C-0021
073.D-0578
192.C-0224
077.C-0364
073.C-0733
077.D-0720
196.C-1006
072.C-0513
074.C-0012
076.C-0878
077.C-0530
078.C-0833
079.C-0681
089.C-0497
183.D-0729
093.C-0919
097.C-0090
60.A-9700
075.C-0332
081.C-0388
072.D-0707
091.C-0853
185.D-0056
081.C-0842
073.C-0784
073.D-0590
089.C-0415
089.C-0732
091.C-0034
092.C-0721
093.C-0409
095.C-0551
097.C-0571
076.C-0155
072.C-0096
073.D-0038
074.D-0131
075.D-0194
076.D-0130
078.D-0071
079.D-0075
080.D-0086
081.D-0065
089.C-0739
074.C-0037
082.C-0427
082.C-0390
096.C-0708

\end{acknowledgements}

\bibliographystyle{aa}
\bibliography{bib42120}

\begin{appendix}


\section{Stellar parameters and correlations}

\begin{landscape}
\begin{table}
\caption{Stellar  parameters and correlations}
\label{tab_sample}
\begin{center}
\renewcommand{\footnoterule}{}  
\begin{tabular}{llllllllllllll}
\hline
Name  & B-V & TS  & T$_{\rm eff}$   & FeH & Number  &  Time  &  $\log R'_{HK}$  &   $\log I_{H\alpha}$ & Global & Nb. of &  LT & Nb. of & ST \\ 
   &  &  &    (K)    &   &  of nights  &  span (nights) &   &  & Correlation & seasons & Correlation & subsets & Correlation  \\ \hline
HD132052 &  0.32 & F0V & 6927$^5$ & -0.01$^5$ &   16 & 1955 &  -4.39 &  -0.59 & -0.02$\pm$ 0.10 &  1 & - &  0 & -\\
HD218396 &  0.26 & F0V & 7656$^6$ & -0.47$^7$ &   15 & 3667 &  -4.19 &  -0.74 &  0.67$\pm$ 0.03 &  1 & - &  0 & -\\
HD109085 &  0.38 & F2V & 6826$^4$ & -0.07$^4$ &   18 & 1957 &  -4.52 &  -0.82 &  0.59$\pm$ 0.10 &  1 & - &  0 & -\\
HD164259 &  0.36 & F2V & 6627$^5$ & -0.19$^5$ &   14 & 2608 &  -4.49 &  -0.76 &  0.08$\pm$ 0.15 &  1 & - &  0 & -\\
HD49933 &  0.35 & F3V & 6535$^3$ & -0.40$^3$ &   26 & 2589 &  -4.49 &  -0.94 & 0.964$\pm$0.004 &  3 & - &  2 &  0.93$\pm$ 0.01\\
HD4247 &  0.30 & F3V & 6945$^4$ & -0.37$^4$ &   17 & 2163 &  -4.32 &  -0.78 &  0.61$\pm$ 0.08 &  1 & - &  1 &  0.17$\pm$ 0.20\\
\hline
\end{tabular}
\end{center}
\tablefoot{The full table is available at the CDS. Only a few lines are shown here to illustrate the content. B, V, and spectral type are from the CDS. Luminosity classes are from the CDS, except for HD163441 \cite[][]{lorenzo18}, HD78534 \cite[][]{dossantos16}, and HD141943 \cite[][]{grandjean20}. T$_{\rm eff}$ values have been shifted depending on the reference to produce a  data set that is as homogeneous as possible following the temperatures provided in \cite{sousa08}, as in \cite{meunier17b}. The number of seasons refers to those with at least five nights. References for T$_{\rm eff}$ and metallicities, indicated as exponents, are the following: (1) \cite{sousa08}, (2) \cite{ramirez14}, T$_{\rm eff}$ shifted by +5 K, (3) \cite{marsden14}, T$_{\rm eff}$ shifted by +13 K (4) \cite{gray06}, T$_{\rm eff}$ shifted by +42 K, (5) \cite{holmberg09}, T$_{\rm eff}$ shifted by -71 K (6) \cite{ammons06}, T$_{\rm eff}$ shifted by +61 K, (7) \cite{gaspar16}, (8) \cite{mortier13}, (9) \cite{soubiran16}, and T$_{\rm eff}$ shifted by +16 K. 
}
\end{table}
\end{landscape}


\section{Correction of H$\alpha$ dependence on B-V}

\begin{figure}
\includegraphics{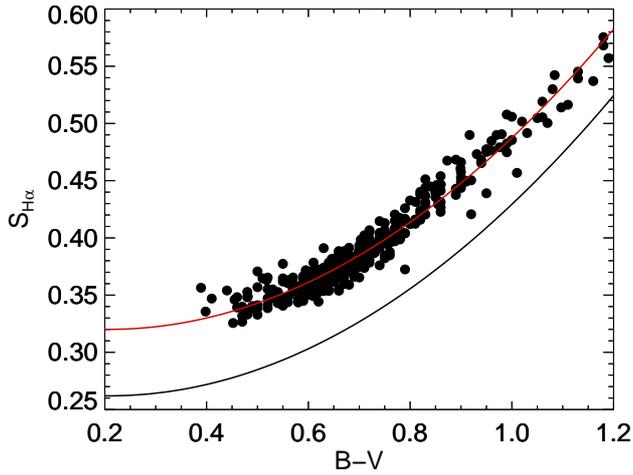}
\caption{
S$_{H\alpha}$ vs. B-V for stars with $\log R'_{HK}<$-4.6. The red  line is the polynomial fit. The black solid line is the same curve shifted downwards to provide positive indexes after correction. 
}
\label{corr_halpha}
\end{figure}

Following the method used by \cite{gomes14}, we fitted the average S$_{\rm H\alpha}$ versus B-V for stars that are not particularly active ($\log R'_{HK}$ below -4.6). We then shifted the curve so that all values are positive  after subtraction of the function, i.e.:

\begin{equation}
\log I_{H\alpha} = \log ( S_{\rm H\alpha}-(0.2733-0.1098\times (B-V)+0.2659 \times (B-V)^2) )
.\end{equation}

The law is illustrated in Fig.~\ref{corr_halpha}. The offset is somewhat arbitrary, as our main focus is on the variability of the Ca II and H$\alpha$ emission and not on the absolute value of the emission.

\section{Synthetic H$\alpha$ time-series due to plages}

We reconstructed H$\alpha$ time-series from Ca II time-series, based on the assumption that only plages are contributing. These were then used to compute the correlation or the variability amplitude, whose properties were compared to the observed ones. Here we describe the three assumptions (Appendixes C.1 to C.3) we considered. We then describe how the H$\alpha$ time-series were built based on these laws, first from the observed Ca II time-series, and then from synthetic Ca II time-series.

\subsection{Ca versus H$\alpha$ emission slope expected from plages for active stars: Hypothesis A}

\begin{figure}
\includegraphics{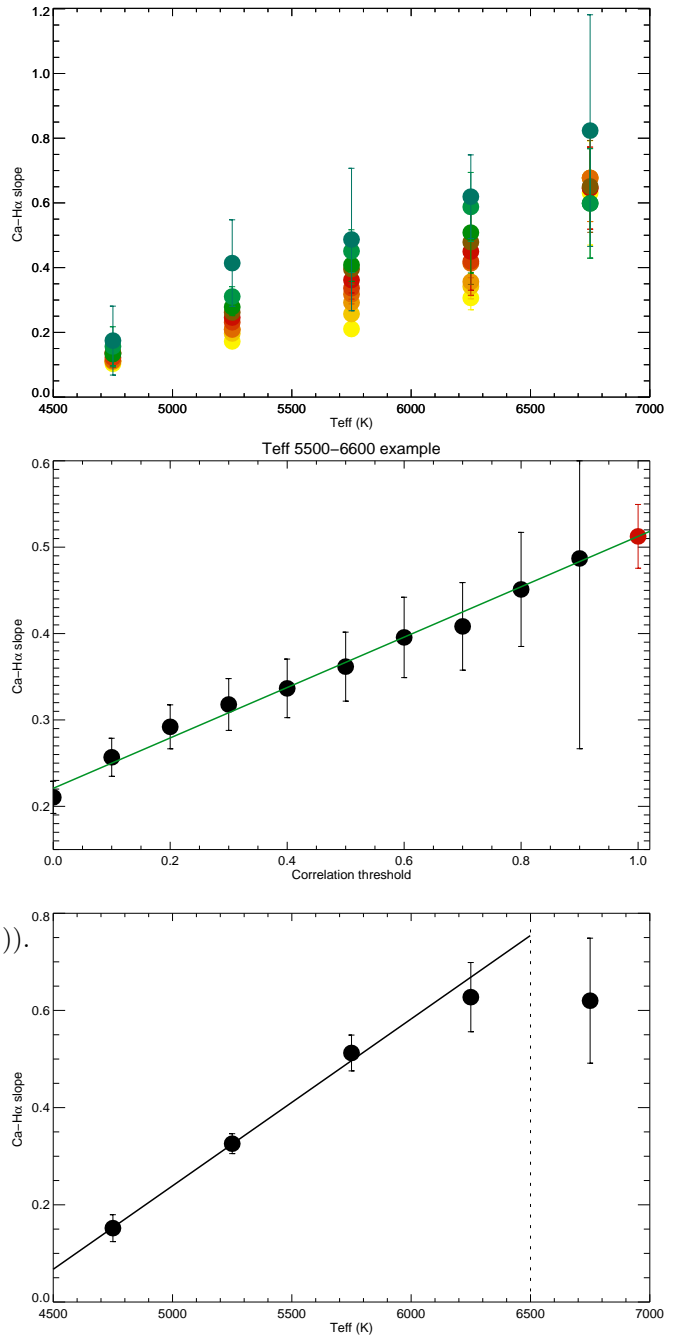}
\caption{
{\it Upper panel:}  Slope between S$_{\rm H\alpha}$ and S$_{\rm Ca}$ binned in T$_{\rm eff}$, for different thresholds in correlation between 0 (yellow) and 0.9 (blue). 
{\it Middle panel:} Slope versus correlation threshold for the bin 5500-6000 K, in black. The red dot corresponds to the extrapolation at a correlation of 1. 
{\it Lower panel:} Extrapolated slope for a correlation of 1 versus T$_{\rm eff}$. 
}
\label{slope}
\end{figure}

There is a possibility that plages in stars of different activity levels follow the same Ca II-H$\alpha$ relation, while the correlation properties and dispersion in H$\alpha$ may be due to other effects (filaments and/or basal flux, for example). In this case, we assumed that the slope we observe between Ca II and H$\alpha$ time-series for active stars with a very good correlation between the two indexes can be used for other stars with similar T$_{\rm eff}$ (Hypothesis A), i.e. a behaviour similar to the Sun (or more generally stars with a very good correlation between the two indexes), for which a correlation of 0.8 was observed \cite[][]{meunier09a}. We defined this slope as follows. For each T$_{\rm eff}$ bins (500 K), we selected stars with a global correlation above a certain threshold and computed the average slope H$\alpha$ versus Ca II for each star. These slopes are shown in the upper panel of Fig.~\ref{slope}. We then considered those average slopes as a function of the threshold in each T$_{\rm eff}$ bin (an example for T$_{\rm eff}$ in the 5500--6000K range is shown in the middle panel) and extrapolated this slope to a correlation of 1 (red point): this corresponds to a situation with a perfect correlation, which would be due to the presence of plages alone, with similar properties for all stars. This final slope is shown versus T$_{\rm eff}$ in the lower panel. There is a clear linear trend for the first four bins, and in the following, we describe the Ca II-H$\alpha$ slope using the function -1.47812+0.000343441$\times$T$_{\rm eff}$. For stars with a   T$_{\rm eff}$ higher than 6500 K, we used the value obtained for the last bin. The uncertainty on the slope is considered when building the synthetic time-series (see below).

\begin{figure*}
\includegraphics{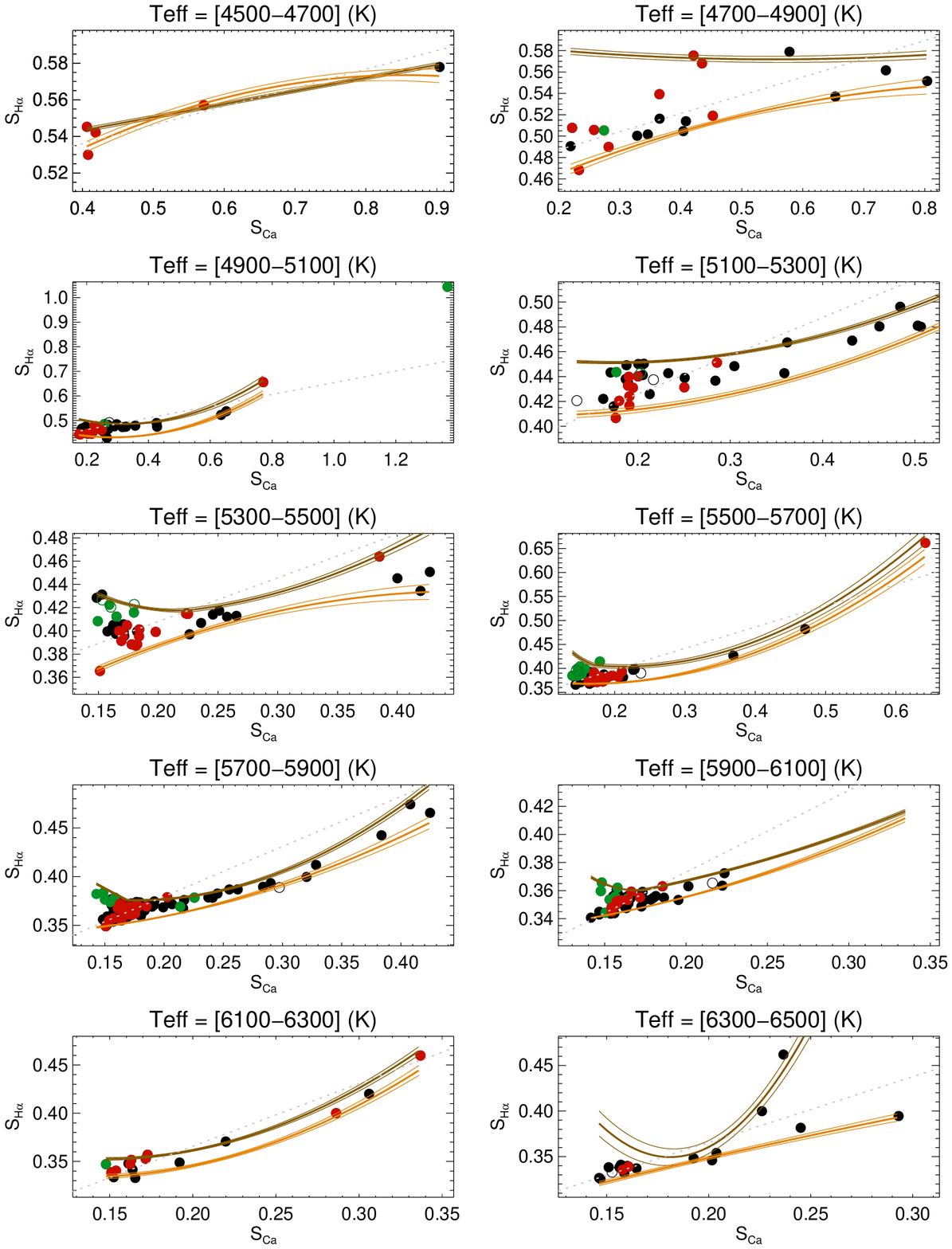}
\caption{
Same as Fig.~\ref{moyzoom} but for the whole range of activity levels. The curves in orange (lower bound) and brown (upper bound) are used to define the laws relating the H$\alpha$ emission in plage to the Ca II emission in hypothesis B. The thick lines are the main laws used in the analysis, while the two thin lines of each colour correspond to the uncertainty. 
}
\label{moybte}
\end{figure*}

\subsection{H$\alpha$ versus Ca II slope from average emission index relations: Hypothesis B}

The second assumption allows for different correlations between the Ca II and H$\alpha$ emission depending on stars, and in particular on their average activity level. The second assumption we considered therefore relies on how  the average H$\alpha$ emission relates to Ca II, studied in Sect.~3. 
As in Sect.~3, we analysed bins in T$_{\rm eff}$   separately to avoid spectral type dependence. In each bin, we determined a lower bound and a upper bound for the H$\alpha$ versus Ca II relation,  corresponding to two extreme configurations, shown in brown and orange in Fig.~\ref{moybte}. For a given star, with a certain average S$_{\rm Ca}$, we considered its S$_{\rm H\alpha}$ values and interpolated the slope used to compute H$\alpha$ (defined as the slope of the H$\alpha$-Ca II relation times the Ca index) from the two extreme curves. 

The lower envelope was computed as follows. We performed a Monte Carlo simulation in which we generated a large number of  polynomials of degree 2, and used two criteria: the distance to the observed points (which we wish to minimise), and the number of points $N$ left below the polynomial curve. For realisations corresponding to $N$=0, we considered the minimum distance. Then, we considered all polynomials such that the distance is lower than this value and $N$ lower or equal to two. This provides a set of realisations, over which an average slope (at any given Ca level) is computed, providing the thick orange line in Fig.~\ref{moybte}. The slope of this curve was used to reconstruct H$\alpha$ time-series from Ca II time-series for stars corresponding to this Ca and H$\alpha$ average levels. The rms of the values over the different curves is used to derive an uncertainty range, shown by the thin curves, which is taken into account when building the time-series. 

The upper envelope was computed in a similar way for most temperature bins. There are four bins (5300--6100 K) for which the variability of the envelope is much better fitted with two polynomials, one corresponding to low-activity stars and the other to the higher activity stars. The resulting curves are shown in brown in Fig.~\ref{moybte}.

\subsection{H$\alpha$ versus Ca II slope from long-term relationship : Hypothesis C}

\begin{figure}
\includegraphics{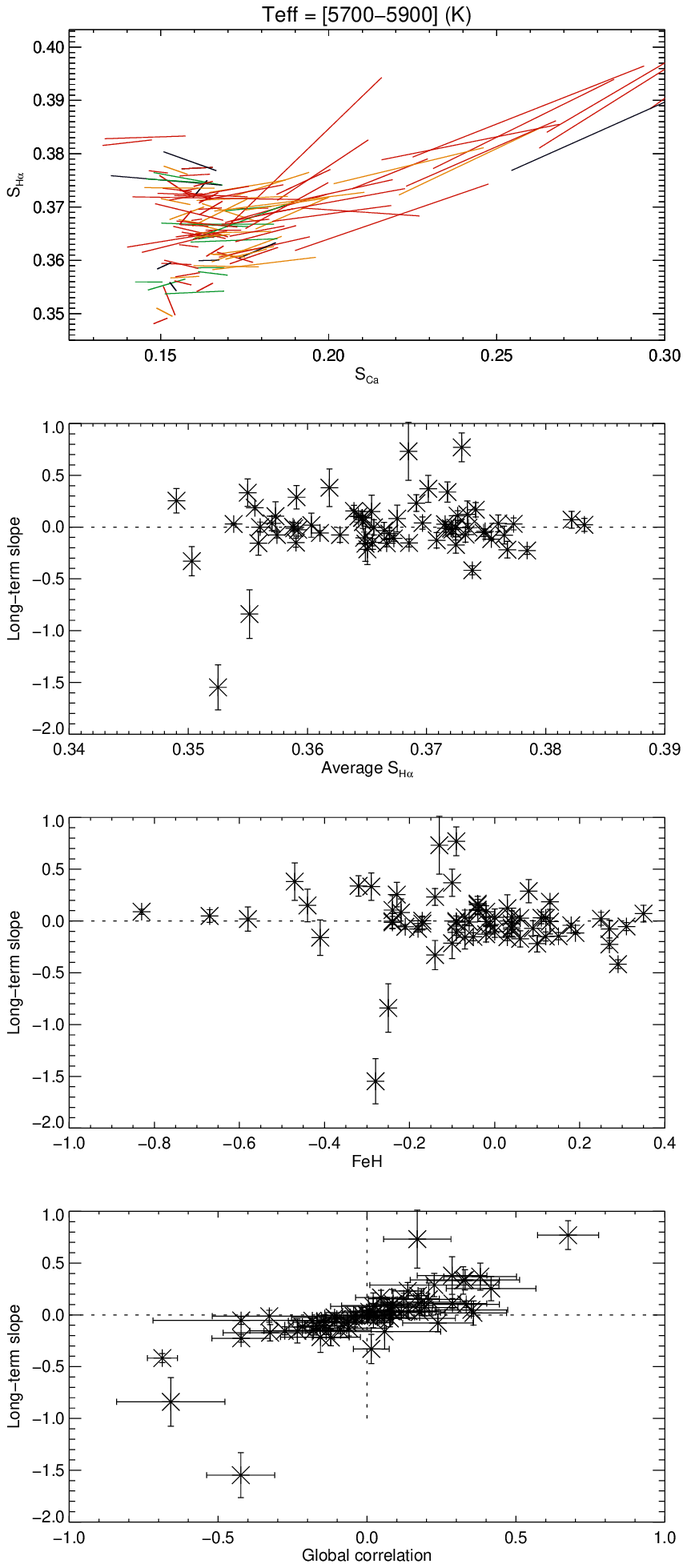}
\caption{
{\it First panel}: Linear fits on S$_{\rm H\alpha}$ versus S$_{\rm Ca}$, for the observed coverage in S$_{\rm Ca}$. The colour code corresponds to the number of nights for each star: between 10 and 20 nights (black), between 20 and 50 nights (red), between 50 and 100 nights (orange), and higher than 100 (green). 
{\it Second panel}: Slope (S$_{\rm H\alpha}$ versus S$_{\rm Ca}$) versus average S$_{\rm H\alpha}$ for quiet stars (S$_{\rm Ca}$ below 0.17) with T$_{\rm eff}$ in the 5700-5900~K range (upper panel). 
{\it Third panel}: Same versus metallicity.
{\it Fourth panel}: Same versus global correlation.
}
\label{hypc}
\end{figure}

Finally, we considered a third model for the H$\alpha$-Ca relation in plages: we assumed that, for a given star, the LT relationship (defined on seasons) is representative of the plage emission in both indexes. The resulting slope was then used to reconstruct synthetic time-series from the Ca time-series (see the following section). We note that for stars for which the  LT  slope is not available, we used the global slope, but this does not concern stars in categories 1 to 3. The uncertainty on the slope is taken into account when building the synthetic time-series.  Figure~\ref{hypc} illustrates the behaviour of the LT  slope for the 5700-5900 K range. The upper panel shows that for active stars (S$_{\rm Ca}$ higher than 0.16), the two indexes are always correlated. The lower panels show those slopes versus the H$\alpha$ average level and metallicity, without an obvious trend. 
If those very different slopes are due to plage only, then there must be other physical processes than activity levels and metallicity affecting the emission in Ca and H$\alpha$ in plages, or there are additional processed in addition to plages to affect this slope. With this assumption, we therefore expect that the LT  variability in H$\alpha$ is well reproduced, and simply seek to know whether or not the ST variability is well retrieved.

\subsection{From observed Ca II time-series to reconstructed H$_\alpha$}

To reconstruct H$\alpha$ time-series from Ca II observations, we proceeded as follows. For hypotheses A and C, we used the slope determined in Appendix C.1 and C.3 respectively, which depends only on the temperature of star (A) or that of star (C). For hypothesis B, we identified the position of the star in the average Ca II-H$\alpha$ diagram, and identified the upper and lower values of the average level and slope corresponding to their average S$_{\rm Ca}$. The synthetic H$\alpha$ was defined as this slope multiplied by the Ca II time-series, and an offset was applied to obtain an average H$\alpha$ level similar to the observed one. White noise was then added using the uncertainty on the H$\alpha$ measurements at each time. This leads to three synthetic H$\alpha$ time-series (A, B, and C).

In addition, we built two additional time-series for each model, corresponding to a lower bound and upper bound at the 1-$\sigma$ level. This was done by considering the uncertainty on the slopes as defined in Appendixes C.1 to C.3. 

\subsection{From synthetic  Ca II time-series to reconstructed H$_\alpha$}

We first constituted a sample of synthetic time-series  similar to the observed ones from the simulations performed in \cite{meunier19}. Those simulations of complex activity patterns are designed to reproduce a realistic behaviour of F-G-K old main sequence stars, more specifically in the F6-K4 range. The main advantages conferred by this model are a better statistical significance as several samples can be generated, better noise effect estimation as Ca II time-series are noise-free, and eventually the possibility to look at better temporal sampling. These simulations take  the impact of inclination into account, in particular in the construction of the chromospheric emission time-series: \cite{meunier19} showed that inclination has an impact on the average level \cite[also found later by][]{sowmya21}, and also impacts the amplitude of the LT  variability. These effects are therefore included in our reconstructions. 

\begin{itemize}
\item{Simulations which are compatible with the star were selected: we used simulations obtained for the closest B-V values, for a random inclination (using a random distribution in cos(i)). We then  selected those with the closest average $\log R'_{HK}$ value and the closest dispersion in S$_{\rm Ca}$. }
\item{A simulation was chosen randomly in that selection, and  the observed temporal sampling for that star was applied. The series was then normalised to provide the same dispersion as the observed one, because the simulations do not cover all values. The S-index was used to build the synthetic time-series as described in Appendix C.4: the only difference is that in this case there is no noise on the initial Ca II time-series used to build the synthetic H$\alpha$ time-series.  }
\item{Random noise was independently added to Ca II and H$\alpha$ time-series using the measured uncertainties for that star. }
\end{itemize}

We note that in principle there is an intrinsic variability from one plage to another for a given star, as observed for the Sun. The present simulations take into account the variability in Ca II emission depending on their size following \cite{harvey99}. Although these latter authors observed a dispersion in the relationship between Ca II emission and magnetic flux for a given category of plages (suggesting that other ingredients control the Ca emission),  we considered here a single relation between the Ca II and H$\alpha$ emission for all plages of a given star for simplicity. \cite{labonte86b} and \cite{robinson90} showed that there is a dispersion in this relation as well (e.g. that observed for the Sun), which is not taken into account here, as this is beyond the scope of the present paper.


\section{Model with filaments}

\subsection{Long-term models: cycle modulation}

\begin{figure}
\includegraphics{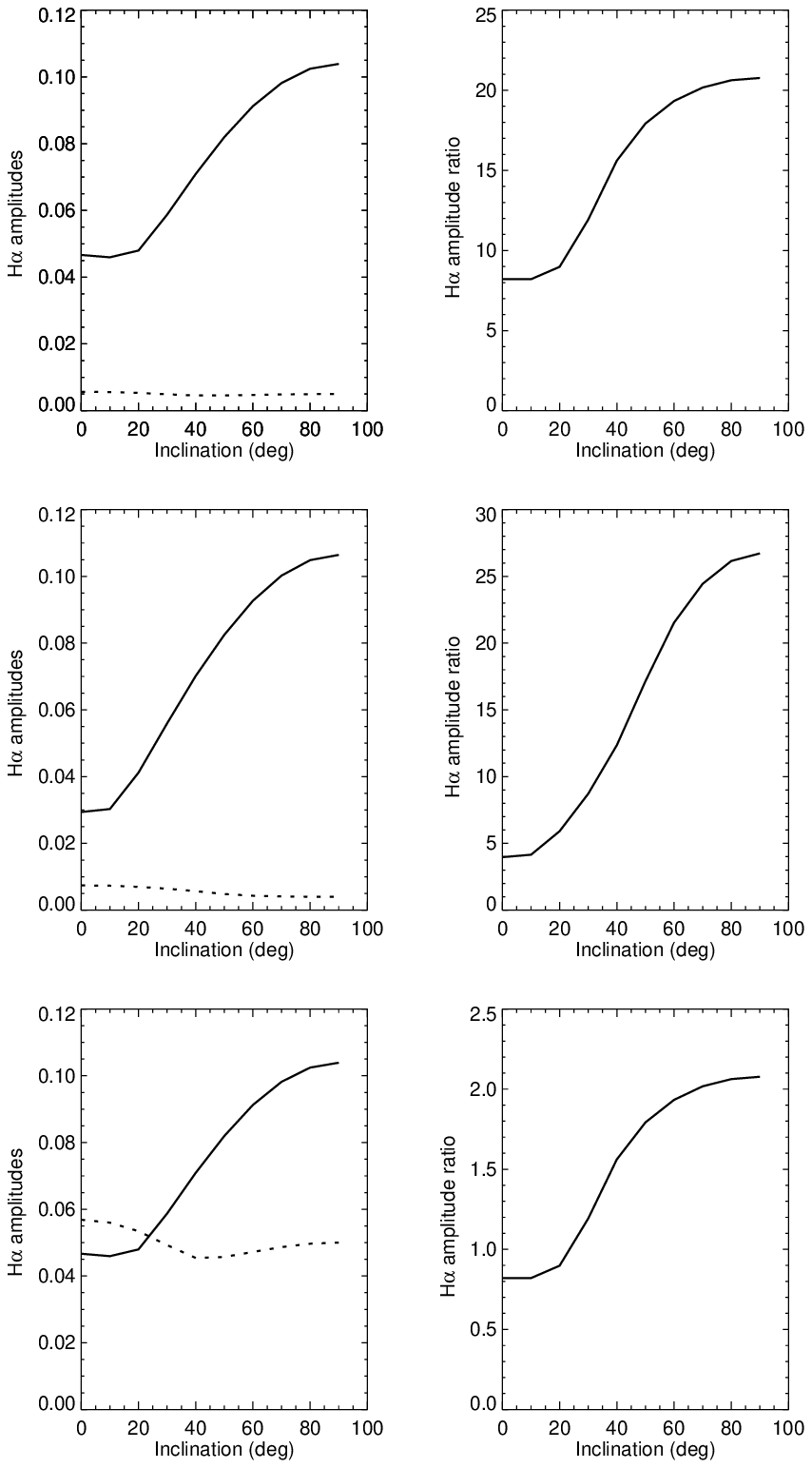}
\caption{
H$\alpha$ amplitude (right panels) versus inclination due to plages (solid line) and filaments (dashed line), and ratio between plage and filament amplitudes (left panels) for different configurations: average solar latitude (LT1 in Table~\ref{tab_config_lt}) and $r$=10 (upper panels), stronger latitude differences (LT2) and $r$=10 (middle panels), and average solar latitude and $r$=1 (lower panels). 
}
\label{ex_lt}
\end{figure}

In order to consider the  LT  contribution of plages and filaments, we considered a model based on latitude bands, in which there is a certain filling factor of each type of structure. In this case, we did not need to take their longitude distribution into account. Plages and filaments were then modelled in two latitude bands, which are symmetric with respect to the equator.  The latitude was either fixed over time, or followed a butterfly diagram trend (Fig.~\ref{butterfly}).  The filling factor was modulated by a solar-like cyclic variation, corresponding to the shape of cycle 23 for sunspots. The apparent filling factor then depended on stellar inclination, between 0$^\circ$ (poleward) and 90$^\circ$ (equatorward). At each time step, we reconstructed the H$\alpha$ emission according to:

\begin{equation}
E_{H\alpha}(t)=C_{\rm pl} \times {\rm ff}_{\rm pl}(t) - C_{\rm fil} \times {\rm ff}_{\rm fil}(t)
,\end{equation}

where the contrasts are the solar values used in \cite{meunier09a}, i.e. 0.35 for plages and 0.2 for filaments. The filament filling factor was typically $r$ times lower than the plage filling factor. We used the Sun as a reference, with a typical ratio $r$ of 10 between the coverage between plage and filaments. The two components could be used to compute separately the amplitudes due to plage and filament. The difference between the maximum and minimum of the plage contribution was first computed; the difference between the filament contributions at the same times provides the filament amplitude.   
A few examples are shown in Fig.~\ref{ex_lt}. In the solar configuration, the relative filament amplitude is larger for a star seen pole-on, with a gain of a factor $\sim$2 compared to edge-on. However, this is not sufficient for  filaments to compensate for the plage contribution, because the ratio (right panel) is around 8, which is significantly larger than 1. A larger difference in latitude between plages and filaments allows to decrease this ratio (obtaining a factor 4 instead of 8 in this example). However, filaments with a ten times larger  contribution than the solar ones ($r$=1 lower panels) would be sufficient to provide an anti-correlation for stars seen close to pole-on. A lower $r$ value compared to the Sun could be reached by changing several properties of the filaments: size, contrast, lifetime, and number.

\subsection{Short-term models: rotational modulation}

To study the effect of the filament parameters on the ST behaviour, we built a very simple model including one plage characterised by its latitude, size, and contrast, and one filament characterised by its latitude, difference in longitude with the plage, size, and contrast. Inclinations between edge-on and pole-on are considered for each configuration.  The reference contrasts and formula to build the H$\alpha$ emission are similar to the previous section. The Ca emission was computed using only the plage contribution and a contrast of 0.7 as in \cite{meunier09a}. 
As before, we considered a ratio $r$ of typically 10 between the plage and filament coverages, as for the Sun, and then studied the effect of smaller ratio to simulate a larger contribution of filaments.

\begin{figure}
\includegraphics{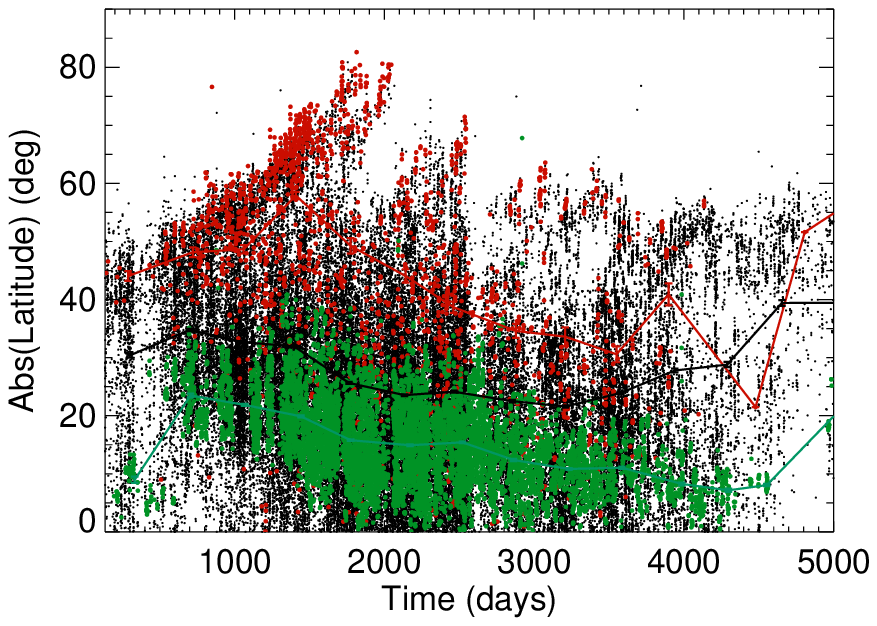}
\caption{
Butterfly diagram (folded between the two hemispheres) of plages (green) from MDI/SOHO, small filaments (black) and large filaments (red) from the BASS2000 data base,  defined with a threshold of 50$^{\circ 2}$. The solid lines correspond to a binning over 1 year. 
}
\label{butterfly}
\end{figure}


\section{Properties and examples from the three categories of stars}

\subsection{Properties of the three categories of stars}

\begin{figure}
\includegraphics{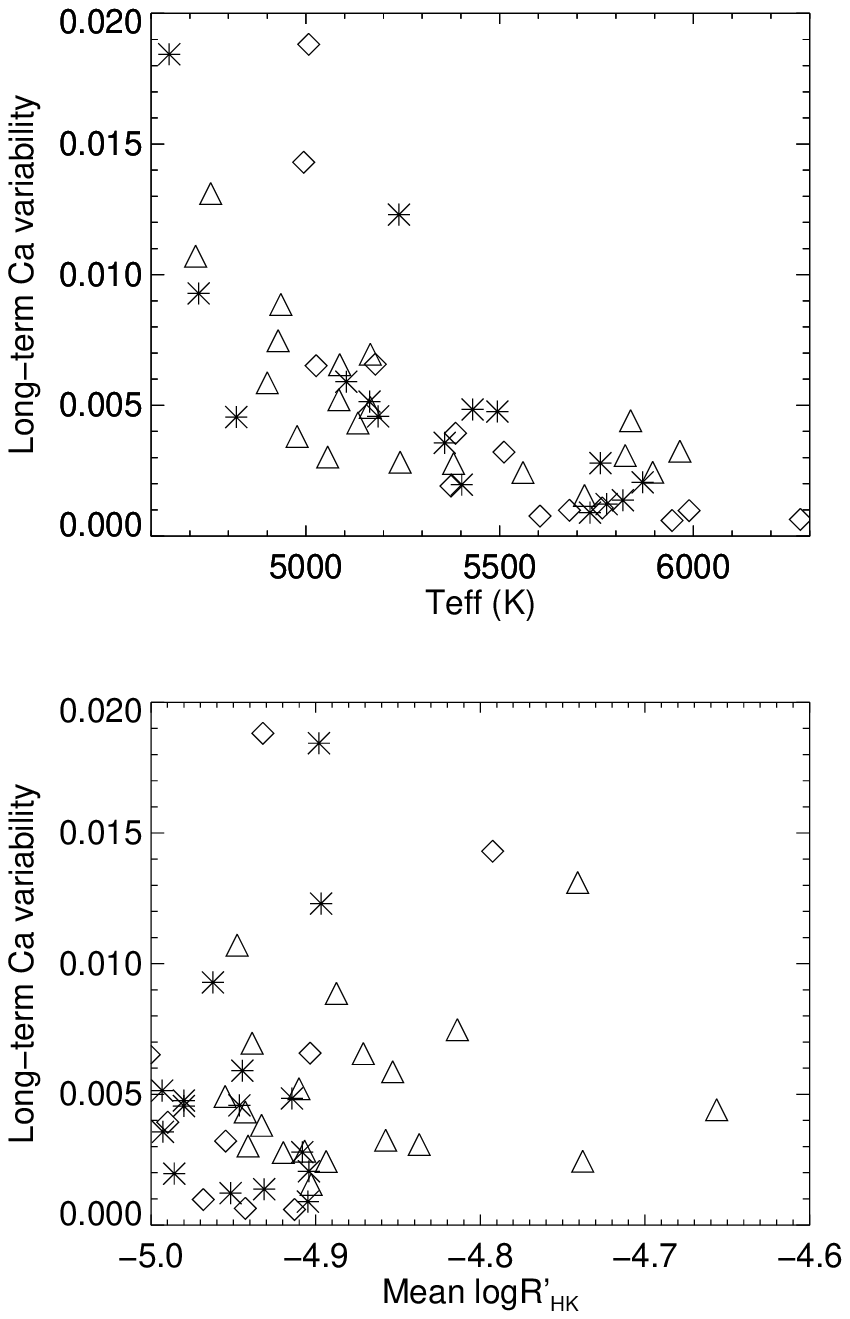}
\caption{
LT  Ca II variability versus T$_{\rm eff}$ (upper panel) and $\log R'_{HK}$ (lower panel) for the three categories: \#1 (stars), \#2 (diamonds), and \#3 (triangle). 
}
\label{ca_categ}
\end{figure}

In this section, we illustrate complementary properties of the stars belonging in one of the three selected categories (defined in Sect.~4.1.1). Figure~\ref{ca_categ} shows the LT  Ca II variability versus T$_{\rm eff}$ and $\log R'_{HK}$. The variability of the S$_{\rm Ca}$ index is larger for low T$_{\rm eff}$ stars, but there are stars from each category at all T$_{\rm eff}$. Stars from categories \#1 and \#2 are primarily seen for stars similar to the Sun or less active, while there are more active stars in category \#3. They are discussed in Sect.~4.1.1 and 4.5 respectively. 

\subsection{Time-series from the three categories}

In this section we show a few examples of our H$\alpha$ time-series and the comparison with our three models. For each example (4 in category \#1, 3 in category \#2, and 4 in category \#3, figures \ref{ex_hamod_1} to \ref{ex_hamod_11}), we show the full time-series, on which the LT  and ST rms are provided, followed by a  close-up on one or two seasons to illustrate different configurations.

\subsubsection{Examples from category \#1}

\paragraph{HD146233}

This star is our primary example shown in Fig.~\ref{exemple} for category \#1. Hypothesis A leads to too large LT rms while the ST rms is in good agreement. Hypotheses B and C lead to too small LT rms, while the ST rms are slightly too small. This shows that  there is no adequate slope for the Ca II - H$\alpha$ relation in plages to provide a good agreement for both LT and ST rms for this star. Furthermore, the variability does not appear to be caused by flares, with correlated levels between adjacent nights rather suggesting rotational modulation or structure evolution, for example. 

\paragraph{HD38858} 

This star was one example with a very large H$\alpha$/Ca ST ratio in Fig.~\ref{fig_ratio_obs}. Hypothesis A leads to overly large LT rms and overly small ST rms. Both hypotheses B and C lead to very small rms. It is clear that the important ST variability in H$\alpha$ cannot be explained by plages only. 

\paragraph{HD20003}

In this example, hypothesis A again leads to overly large rms, but on the other hand both hypothesis B and C lead to rms in good agreement for both ST and LT. 

\paragraph{HD69830}

This star was one example with a very large H$\alpha$/Ca  ST ratio in Fig.~\ref{fig_ratio_obs}. Hypothesis A leads to overly large LT rms and overly small ST rms, while B and C lead to a much too small rms (both LT and ST). We also note that for one of the seasons shown, the strong dispersion in H$\alpha$ is due to several days with less emission, which is not compatible with flares. It is in better agreement with the presence of a large dark structure, and in relation to rotational modulation or structure evolution.

\subsubsection{Examples from category \#2}

\paragraph{HD7199}

This star is our primary example shown in Fig.~\ref{exemple} for category \#2. For all stars in this category, hypothesis A does not lead to the proper sign (even when the rms is in good agreement with observation). Here, hypothesis B leads to overly large LT rms while the ST rms is too small. However, hypothesis C is in good agreement with observations for both LT and ST rms. 

\paragraph{HD 106116}

This star was one example with a very large H$\alpha$/Ca ST ratio in Fig.~\ref{fig_ratio_obs}. The LT rms for hypothesis A is in good agreement with observations but does not have the proper sign. Hypothesis B provides LT and ST rms which are too small. Hypothesis C leads to a good LT rms, but the ST rms is too small, which is because of the large ST variation in H$\alpha$ which cannot be reproduced here. 

\paragraph{HD65907A}

This star was one example with a very large H$\alpha$/Ca  ST ratio in Fig.~\ref{fig_ratio_obs}. The LT rms for hypothesis A is in good agreement with observations but does not have the proper sign. Hypothesis B provides LT and ST rms which are too small, especially for the LT rms. The same is observed for hypothesis C, although to a lesser extent compared to HD106116.

\subsubsection{Examples from category \#3}

\paragraph{HD215152}

This star is our primary example shown in Fig.~\ref{exemple} for category \#3. Hypothesis A leads to an overly large rms, especially for the LT rms. Hypothesis B leads to good rms but with an anti-correlation instead of a correlation. Hypothesis C leads to a good agreement with observation for both LT and ST rms, although the ST rms is slightly too small. The close-up on the season shows that the ST variability is well reproduced by  hypothesis C.

\paragraph{HD154577}

All three models provide a good agreement for both the LT and ST rms, but the latter is slightly too small. The  close-up on the season also shows a very good agreement with observations. We note that the global correlation is 0.8 for this star, i.e. similar to the Sun.

\paragraph{HD13808}

Hypothesis A provides a too large LT rms, while the ST rms is in good agreement with observations. On the other hand, hypothesis B  provides amplitudes in agreement with observations but with the wrong sign (anti-correlation instead of correlation). Hypothesis C leads to a reasonable agreement, although the ST rms is slightly too small.

\paragraph{HD85512}

Hypothesis A leads to a too large LT rms, while the ST rms is in good agreement. Hypothesis B leads to an overly small variability for both LT and ST rms. Hypothesis C leads to an overly small rms as well, especially for the ST variability. This a typical example showing that even for a star with a relatively good correlation between Ca II and H$\alpha$ (global correlation of 0.60$\pm$0.01), with a well-defined activity cycle, it is not possible to reproduce both LT and CT variation with a single plage model.

\onecolumn

\begin{figure}
\includegraphics{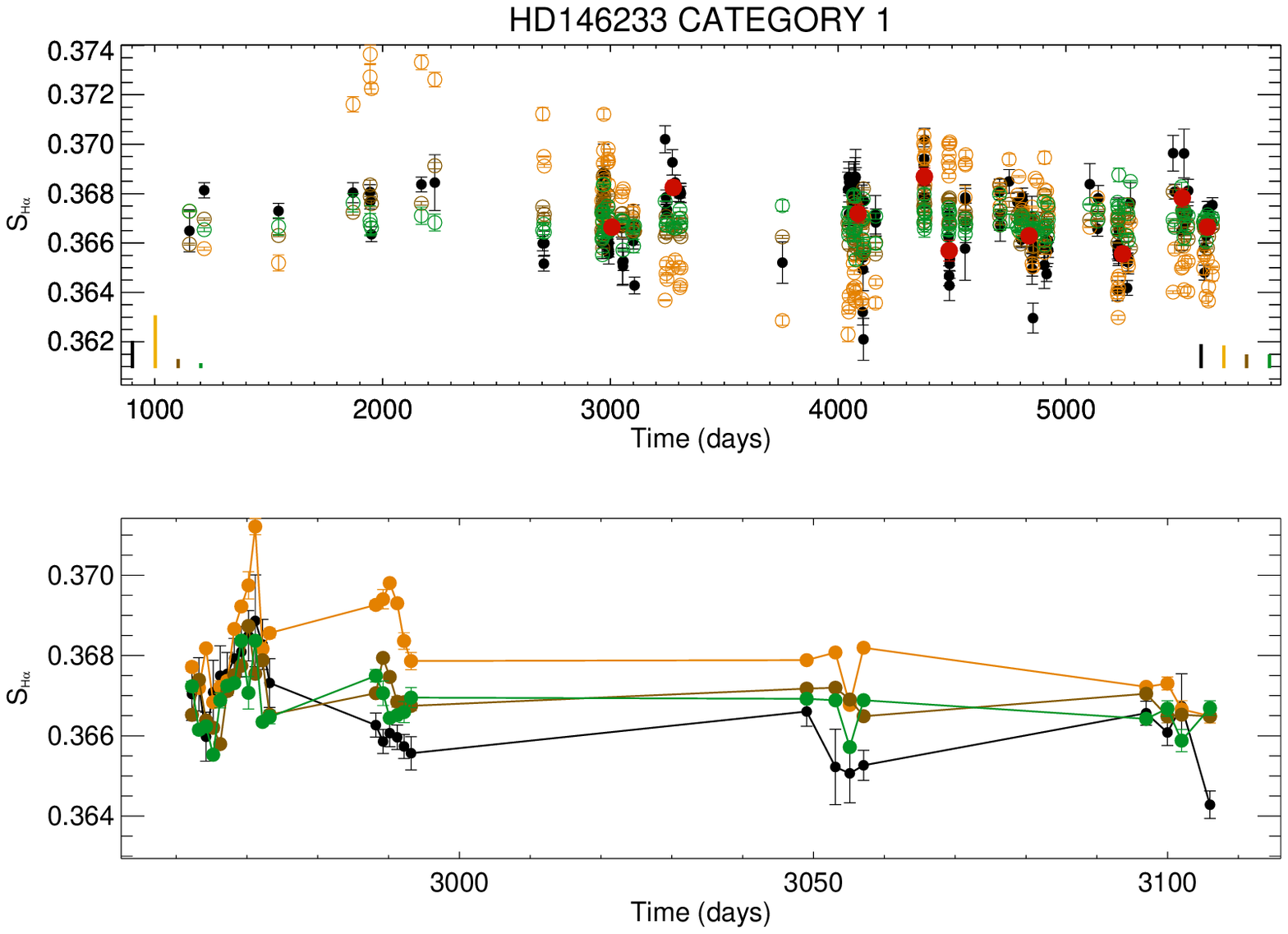}
\caption{
Observed H$\alpha$ time-series (black) compared to model based on Hypothesis A (orange), hypothesis B (brown), and hypothesis C (green) for HD146233 (upper panel). The vertical lines on the left indicate the rms of the LT signal (same colour code) and the vertical lines on the right the rms of the ST signal.  The red symbols correspond to the season averages.  The lower panel shows the variability for one of the seasons (same colour code).  
}
\label{ex_hamod_1}
\end{figure}

\begin{figure}
\includegraphics{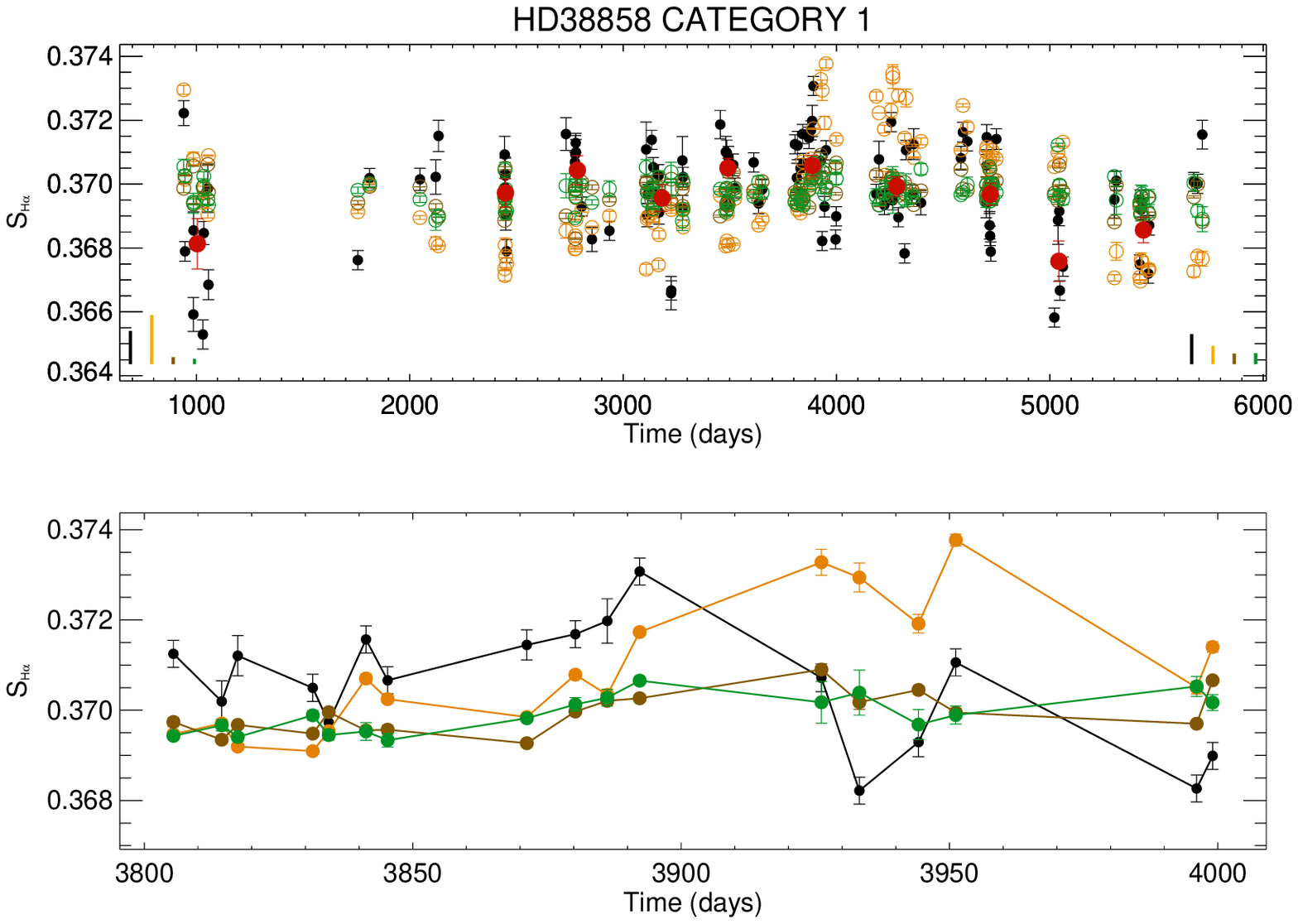}
\caption{Same as Fig.~\ref{ex_hamod_1} for HD38858.}
\label{ex_hamod_2}
\end{figure}

\begin{figure}
\includegraphics{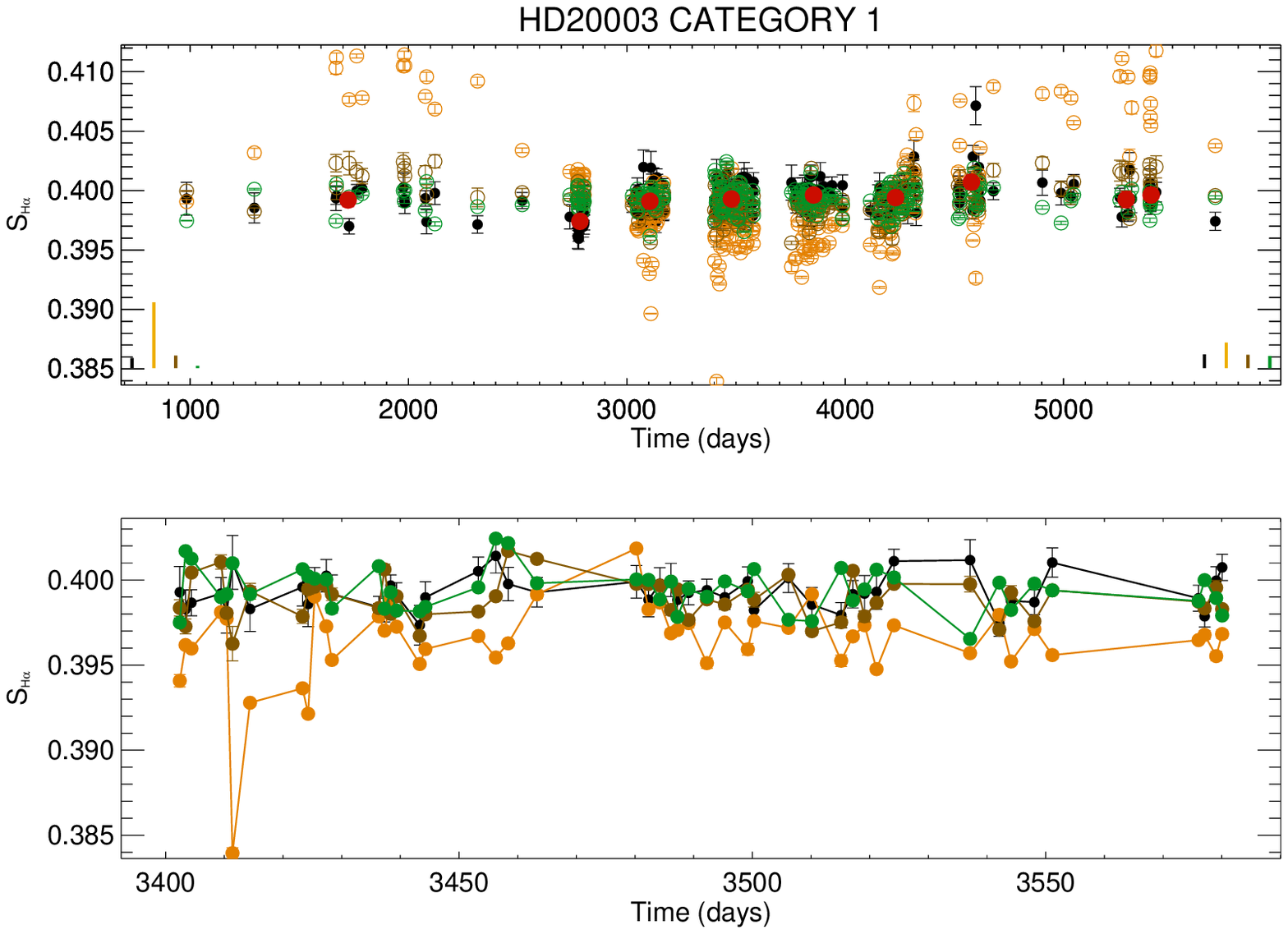}
\caption{Same as Fig.~\ref{ex_hamod_1} for HD20003.}
\label{ex_hamod_3}
\end{figure}

\begin{figure}
\includegraphics{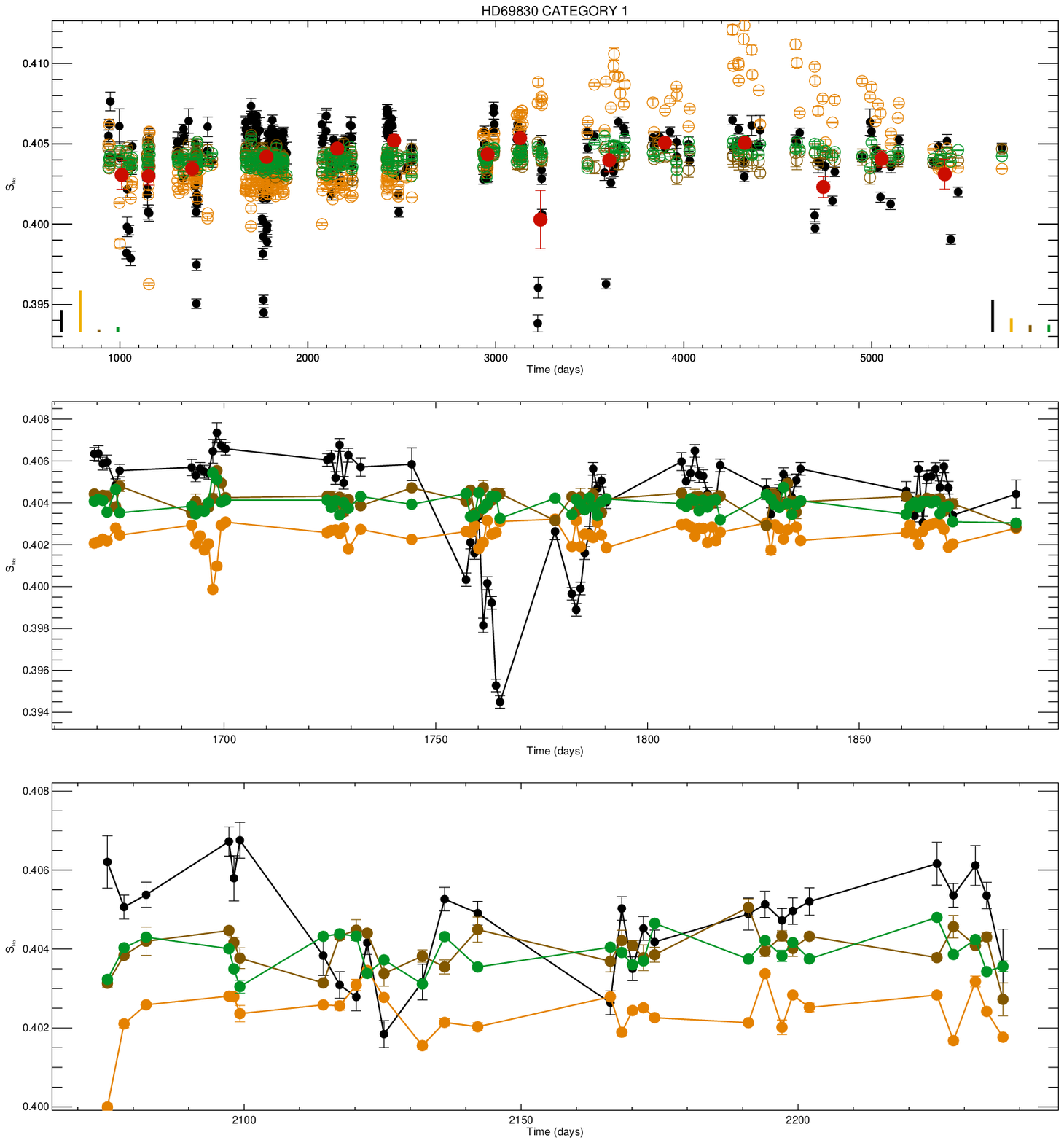}
\caption{Same as Fig.~\ref{ex_hamod_1} for HD69830.}
\label{ex_hamod_4}
\end{figure}

\begin{figure}
\includegraphics{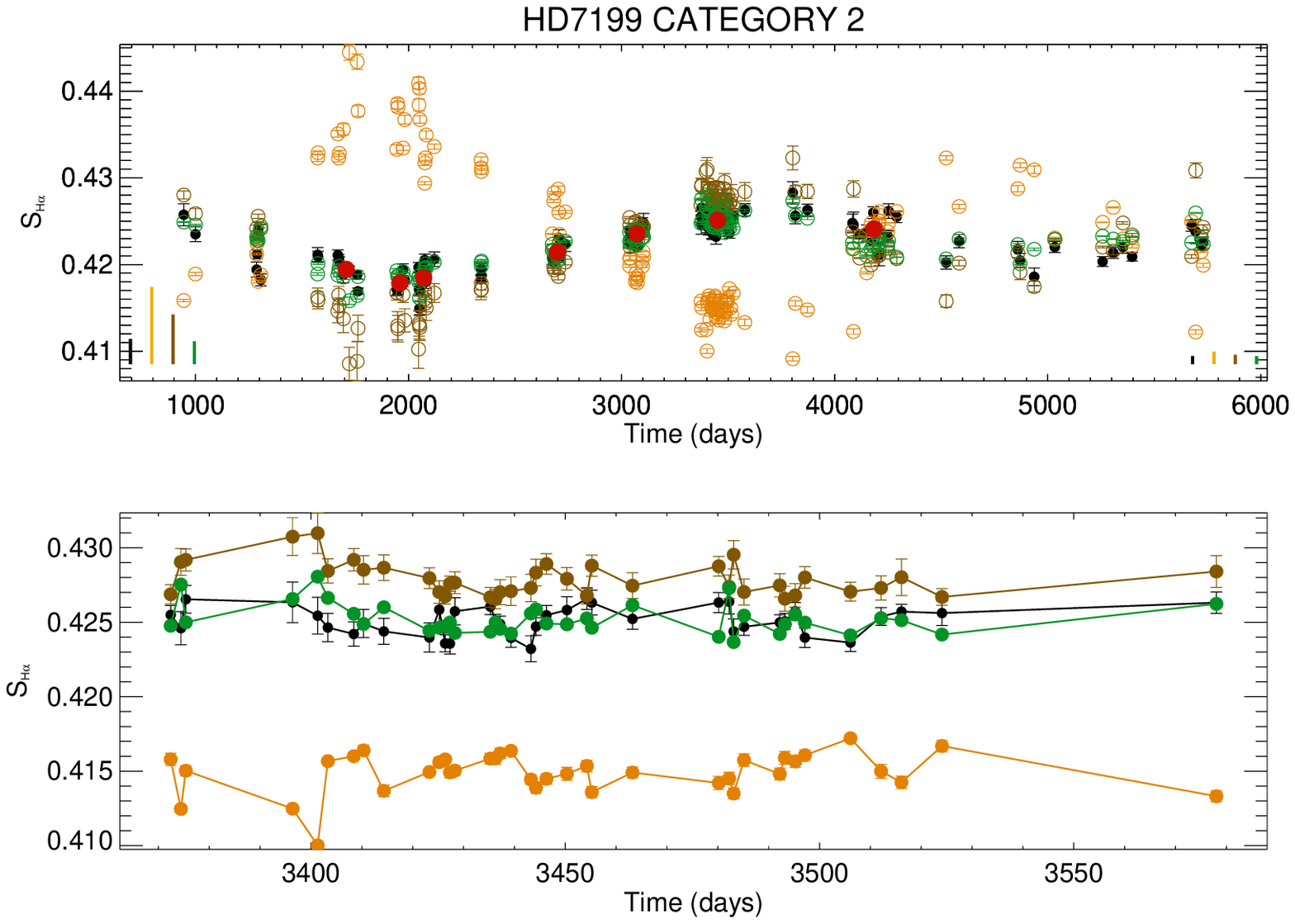}
\caption{Same as Fig.~\ref{ex_hamod_1} for HD7199.}
\label{ex_hamod_5}
\end{figure}

\begin{figure}
\includegraphics{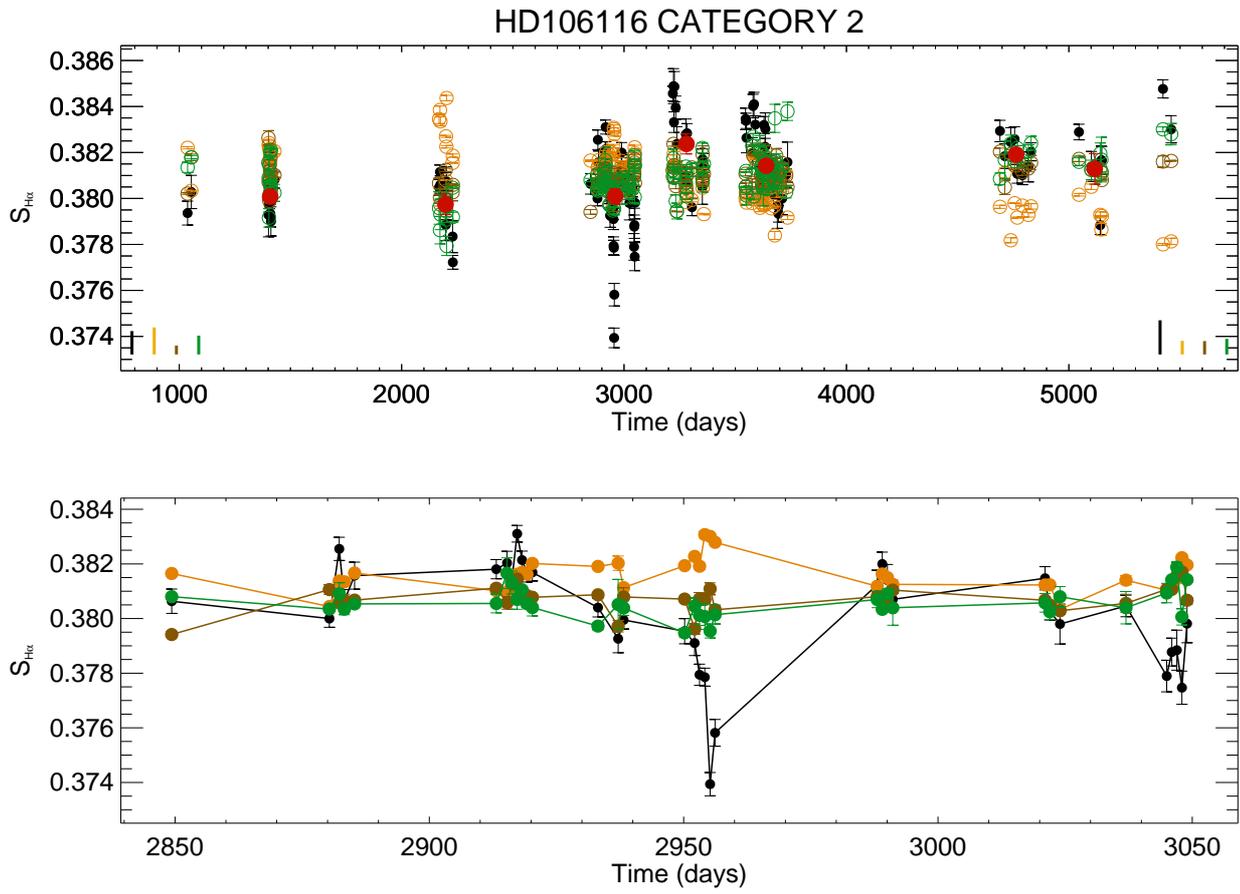}
\caption{Same as Fig.~\ref{ex_hamod_1} for HD106116.}
\label{ex_hamod_6}
\end{figure}

\begin{figure}
\includegraphics{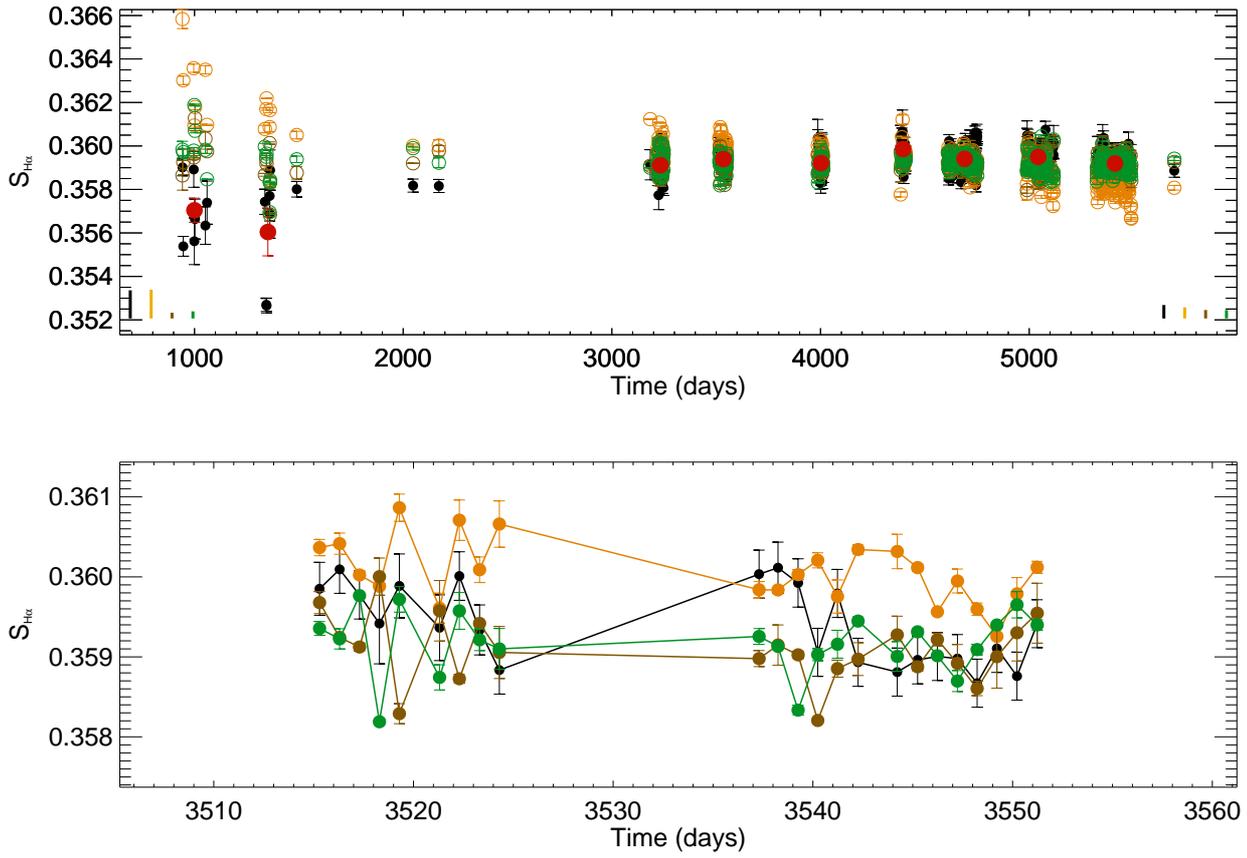}
\caption{Same as Fig.~\ref{ex_hamod_1} for HD65907A.}
\label{ex_hamod_7}
\end{figure}

\begin{figure}
\includegraphics{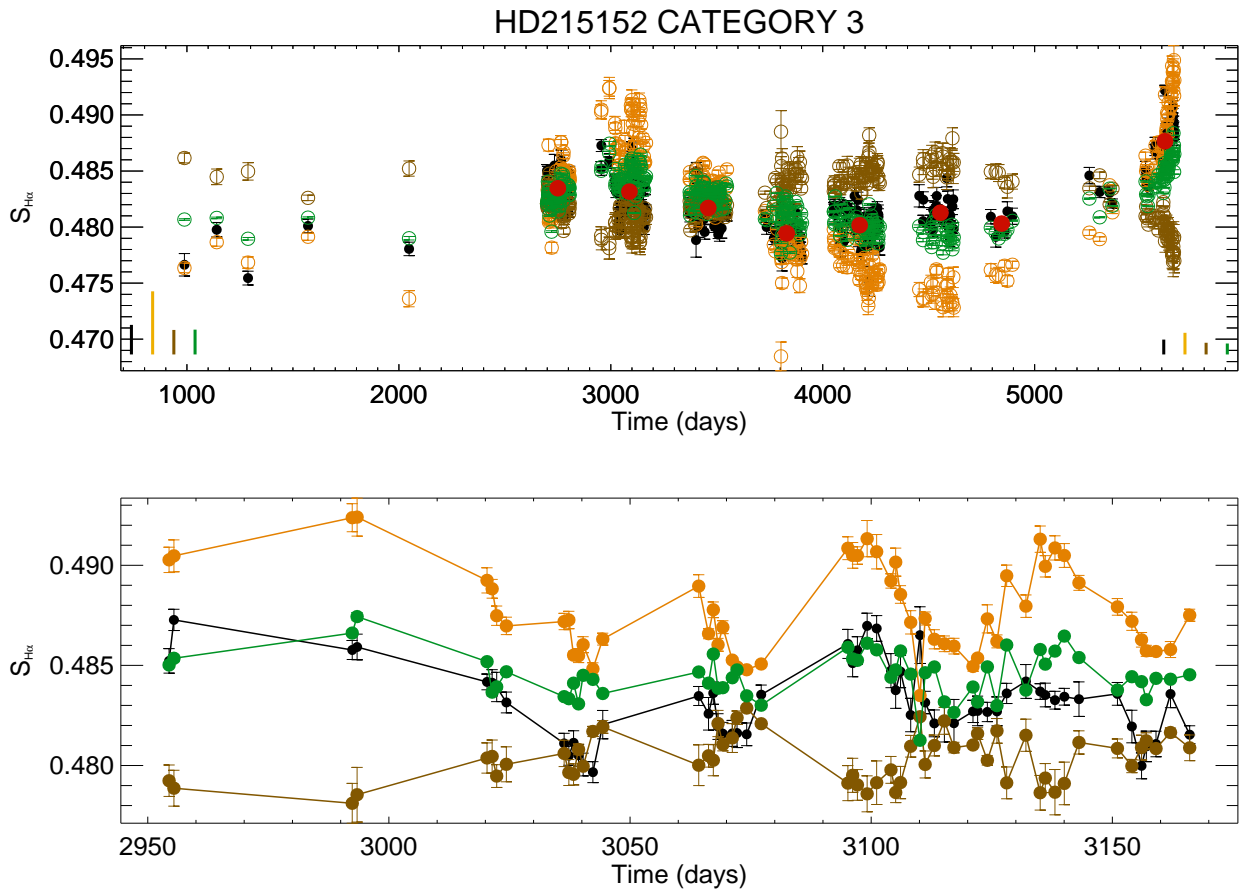}
\caption{Same as Fig.~\ref{ex_hamod_1} for HD215152.}
\label{ex_hamod_8}
\end{figure}

\begin{figure}
\includegraphics{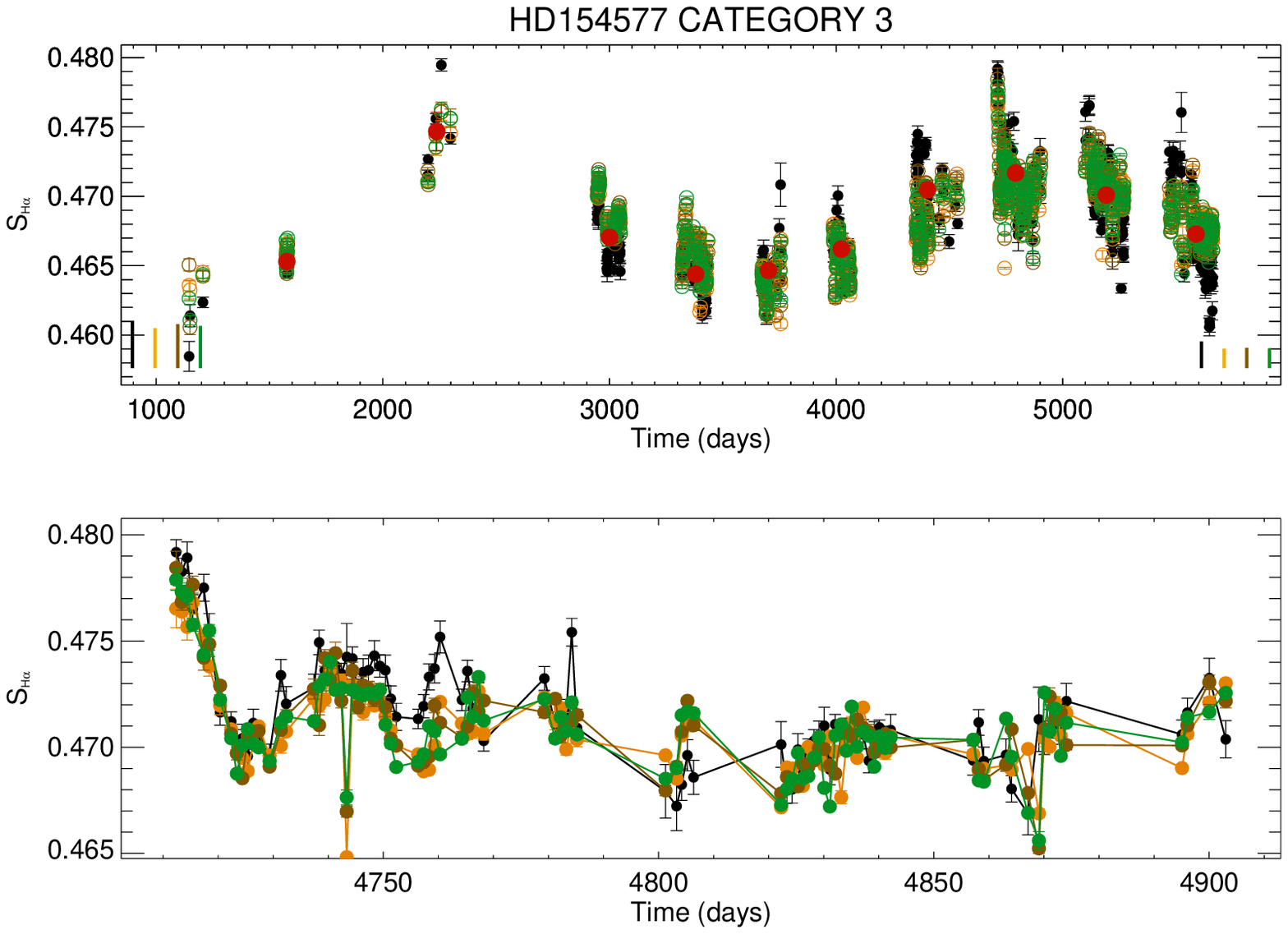}
\caption{Same as Fig.~\ref{ex_hamod_1} for HD154577.}
\label{ex_hamod_9}
\end{figure}

\begin{figure}
\includegraphics{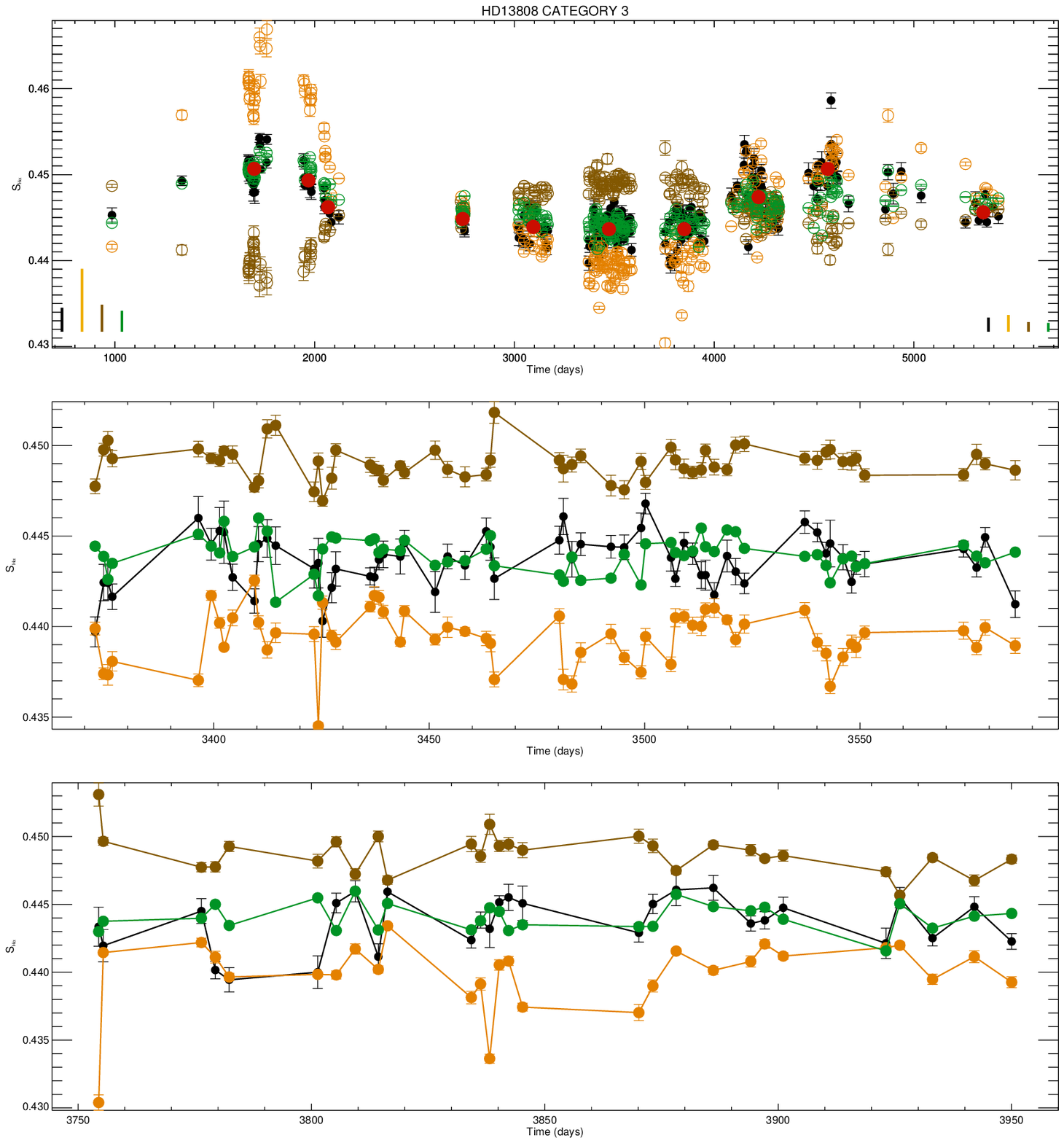}
\caption{Same as Fig.~\ref{ex_hamod_1} for HD13808.}
\label{ex_hamod_10}
\end{figure}

\begin{figure}
\includegraphics{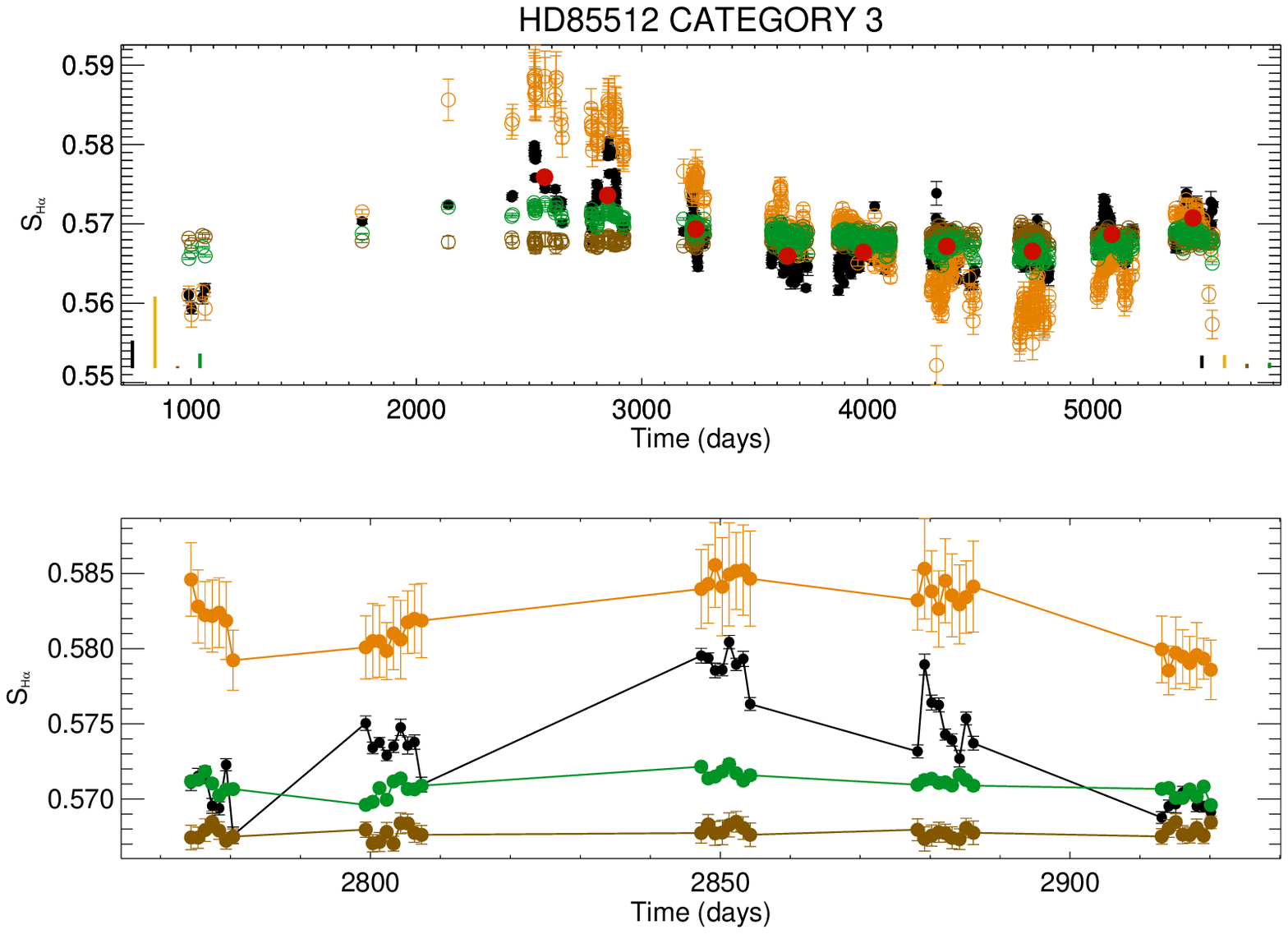}
\caption{Same as Fig.~\ref{ex_hamod_1} for HD85512.}
\label{ex_hamod_11}
\end{figure}

\end{appendix}

\end{document}